\documentclass[aps,twocolumn,noshowkeys,showpacs]{revtex4-1}

\usepackage{dcolumn}
\usepackage{amsmath}
\usepackage{bm}
\usepackage{graphicx}
\usepackage{graphics}

\begin{document}

\title{Sensitive SQUID magnetometry for studying nano-magnetism}

\author{M. Sawicki}
\email{mikes@ifpan.edu.pl}
\author{W. Stefanowicz}
\affiliation{Institute of Physics, Polish Academy of Science, al.~Lotnik\'ow 32/46, PL-02-668 Warszawa, Poland}

\author{A. Ney}
\affiliation{Fachbereich Physik and Center for Nanointegration Duisburg-Essen (CeNIDE), Universit{\"a}t Duisburg-Essen, Lotharstr.\,1, D-47057 Duisburg, Germany}

\date{\today}

\begin{abstract}

The superconducting quantum interference device (SQUID) magnetometer is one of the most sensitive experimental techniques to magnetically characterize samples with high sensitivity. Here we present a detailed discussion of possible artifacts and pitfalls characteristic for commercial SQUID magnetometers. This includes intrinsic artifacts which stem from the inherent design of the magnetometer as well as potential issues due to the user. We provide some guidelines how to avoid and correct these, which is of particular importance when the proper magnetization of nano-scale objects shall be established in cases where its response is dwarfed by that of the substrate it comes with, a situation frequently found in the field of nano-magnetism.

\end{abstract}

\pacs{07.55Jg}

\maketitle

\section{Introduction}
\label{sec: intro}

Many developments in modern solid state and nano-scale physics make very sensitive magnetometry an essential tool to study the magnetic properties of samples like ultrathin films \cite{Ney:2001_EL}, nanoparticles \cite{Sundaresan:2006_PRB}, nanodot arrays or dilute magnetic semiconductors (DMS) \cite{Dietl:2010_NM} which are all deposited on a substrate. Especially in the latter field recent theoretical predictions  \cite{Dietl:2000_S,Sato:2010_RMP} as well as claims of anisotropic ferromagnetism \cite{Carmeli:2003_JCP,Venkatesan:2004_N}, colossal magnetic moments \cite{Dhar:2005_PRL,Dhar:2006_APL} or ferromagnetism in oxides like HfO$_2$ \cite{Venkatesan:2004_N} have triggered research efforts towards novel classes of material with magnetic order at room temperature and theories to explain the observed magnetic behavior  \cite{Coey:2006_COSSMS,Katayama-Yoshida:2008_B}. Nevertheless, there are other studies which discuss the influence of stainless-steel tweezers handling on the magnetic properties of HfO$_2$ \cite{Abraham:2005_APL} and other diamagnets \cite{Garcia:2009_JAP}, point out "possible pitfalls in search of magnetic order" for samples using sapphire substrates \cite{Salzer:2007_JMMM}, or discuss potential contamination of the sample holder itself \cite{Bonanni:2007_PRB}. Most of the exciting claims of unusual magnetic properties have the use of superconducting quantum interference device (SQUID) magnetometry in common. While some work on thin films is done using a home-built ultrahigh vacuum compatible SQUID magnetometer without external magnetic field \cite{Ney:2001_EL}, the majority of the experiments for solid state samples is performed using the commercial SQUID magnetometer MPMS (XL) from Quantum Design \cite{QD}. Using the MPMS SQUIDs is widely spread mainly due to its high degree of user-friendly automation and reliability as well as the lack of commercial alternatives. Ironically, the (apparent) easiness of the MPMS's everyday operation, a virtually effortless sample-to-"*.dat"-file transition, is largely responsible for most of the erroneous reports. Despite the thick user-manual (three thick folders) the highly automated measurement-routines switch off the indispensable critical discussion of the as-received data, an increasing bad practise in science, particularly among the recent breed of scientists who tend to accept computer-yielded results as given and practically error-free. This remark is particularly valid when the proper magnetization of nano-scale objects shall be established in cases where its response is dwarfed
by that of the substrate it comes with.

Here we present a detailed survey of possible artifacts found in measurements using high sensitivity magnetometry, like SQUID magnetometers, with the main emphasis put on the MPMS-SQUIDs, as these have become now a kind of "laboratory standard" in nanoscale magnetism. We firstly discus issues stemming from the measurement technique in general and later proceed to artifacts which originate from the set-up of the measurement by the user. In parallel, we will provide an analysis of typical systematic errors augmented with some recommendation how to avoid those potential research pit-falls.

\section{Commercial SQUID magnetometry}
\label{sec: CommSQUIDmag}

\subsection{General remarks}
\label{sec:GenRemarks}

All measurements have been taken with MPMS XL (5~T) rigs at the University of Duisburg-Essen and Institute of Physics of Polish Academy of Sciences in Warsaw. Nevertheless, the artifacts are visible in other data of low signal measurements recorded with various MPMS machines, e.\,g., on Cr-doped InN \cite{Ney:2007_PRB}, GaAs single crystals \cite{Ney:2006_JPCM} or Gd ion implanted cubic GaN \cite{Lo:2007_APL}, GaN doped with Fe \cite{Bonanni:2007_PRB}, GaN doped with Mn \cite{Stefanowicz:2010_PRBa}. Therefore they can be taken as representative for the entire family of the MPMS instruments. For all measurements the RSO option \footnote{Reciprocating Sample Option measures a sample by moving it rapidly and sinusoidally through the SQUID pickup coil assembly} (4~cm of sample movement and 0.5~Hz to 1~Hz of repetition frequency) of the MPMS was used which offers a higher sensitivity than the standard dc-transport. Whenever required we will use the unit emu, since the absolute size of the total moment $m$, irrespective of the sample volume, determines how close the actual magnetic signals are to the artefact-level of the instrument for low signal samples. The conversion from emu to the SI-unit A/m is straightforward: 1~emu$=10^3$~A/m; the MPMS solely provides $m$ in emu.

\begin{figure}[t]
   \begin{center}
   \includegraphics[width=0.9\columnwidth]{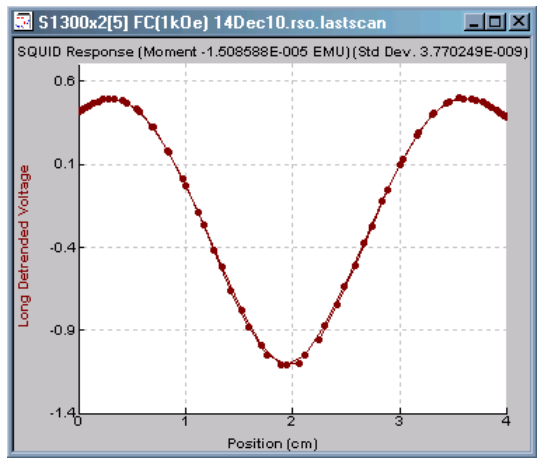}
   \end{center}
   \caption{(Color online) An example of a typical "scan" collected by MPMS system during a single a measurement act using reciprocating sample option mode of measurement. The "Long Scaled Response" is the direct representation of $V(z)$ without any compensation for system drifts applied. Its amplitude corresponds to the strength of the measured moment. In this case the sample is GaN:Mn layer on $5 \times 4 \times 0.6$~mm$^3$ sapphire. This scan was acquired at $H=~1$~kOe and $T=15$~K. }
   \label{fig:ScanQD}
\end{figure}

\begin{figure}[th]
   \begin{center}
        \includegraphics[width=0.99\columnwidth]{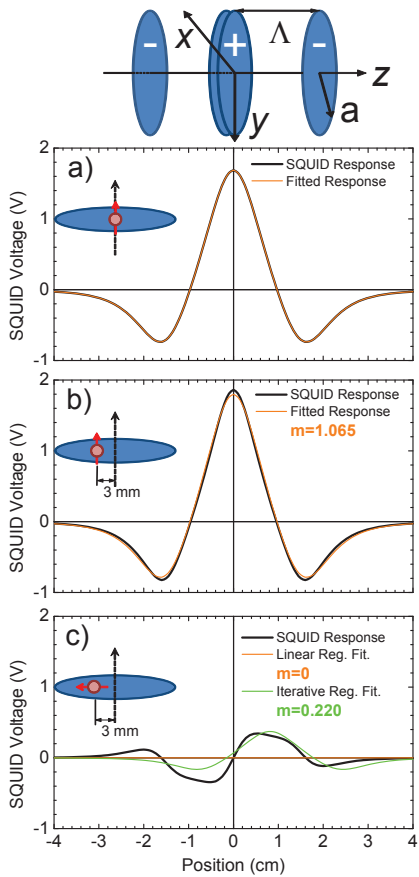}
   \end{center}
   \caption{(Color online) Topmost cartoon sketches the second order gradiometer configuration. In the MPMS system the coil radius $a=$~0.97 cm and the section separation $\Lambda=$~1.519 cm \cite{QD1014-213}. (a) The gradiometer response to a point dipole traveling along $z$ on the axis of symmetry. This is the actual shape of the reference $\Upsilon(z)$ scan that is used to compute sample moment $m$ upon experimentally established scan $V(z)$. (b) Calculated $V(z)$ for the same point dipole as in (a) but 3 mm radially offset from $z$. (c) Same as in (b) but this time the moment of the dipole is located in $xy$ plane. Bold black curves are calculated for the given situation indicated in each of the panels. Thin colored lines are least-squares regression fits of the reference $\Upsilon(z)$. The values of m indicate the relative increase (b) and reduction (c) of the magnetic moment reported by the system at these two experimental configurations.}
   \label{fig:SamplePos}
\end{figure}

Let us begin with a very basic statement that contrary to their common name SQUID magnetometers do not measure directly any magnetic moment as e.g. a moving-iron instrument measures a flow of an electric current. Instead they merely read the current exited in the superconducting loop (called a flux transformer) by a change of a magnetic flux threading the loop, however even this is done in a technically elaborate way. The existence of the flux transformer allows to place the actual SQUID sensor in a protected environment (usually in a niobium shield immersed in the liquid helium bath of the SQUID cryostat), sufficiently away from a sample chamber when the sample can be heated up to material limits (currently 800 K) and the magnetic field can be ramped up almost to the critical field of the superconducting wire the flux transformer is made of. The detecting part of the flux transformer at the sample-end is wound outside a variable temperature insert of the magnetometer, so it primarily transmit rather large changes of the magnetic flux caused by the change of the field in the external magnet (the source of experimental magnetic field) and due to temperature and $H$-field  variations of magnetization of all the components used to make the sample chamber (and actually everything else around). In order to minimize the spurious fluxes of the whole environment and further to separate the minute signal of the specimen from the former, two clever tricks are employed: (i) the sample-end of the flux transformer is shaped into a second order gradiometer, (ii) the sample is transported through the gradiometer sections producing a unique form of a time dependent current, $J_{\Phi}(t)$, in the superconducting flux transformer. Now, as long as time variations of the other sources of the magnetic flux that couple to the flux transformer are sufficiently slower or weaker than the $J_{\Phi}(t)$, a value of the sample magnetic moment $m$ can be established through (most frequently) numerical comparison of expected and measured $J_{\Phi}(t)$, with $m$ being the proportionality factor scaling the former into the latter. We remark here that obviously the sample travel across the detection coils takes place in a well controlled manner, so practically $J_{\Phi}(t)$ is replaced by its dependence on the sample position in the sample chamber $J_{\Phi}(z)$, converted to a "SQUID voltage" $V(z)$ (which is usually called "scan") \footnote{Throughout the MPMS system this is refereed to as an $x$-Position. We adopt here a notion corresponding to Cartesian coordinates since the magnetic field is oriented vertically in lab space (i.\,e. $z$) and the gradiometer has got the same axis of symmetry as the magnet.}. These types of scans are represented graphically by the SQUID software in its user interface (see Fig.~\ref{fig:ScanQD} for the example). This characteristic shape of $V(z)$ reflects the general form of the second order gradiometer, see the cartoon of its actual shape in Fig.~\ref{fig:SamplePos}. The expected gradiometer response function $\Upsilon(z)$ [see Fig.~\ref{fig:SamplePos}(a)] is taken for a {\it point dipole} located on the $z$~axis (the symmetry axis of the gradiometer). It should be noted that the assumption of an ideal point dipole limits the useful sample size to at most 5~mm along the sample travel to keep the systematic error in $m$ below $\sim$3\%.


Here we have to place a strong comment concerning the system sensitivity or resolution for {\it absolute} values of $m$. Both depend on the relevance of the established $V(z)$ to the real experimental configuration and a fidelity of $V(z)\rightarrow m$ numerical conversion, so great care should be taken by the user on every step of the experimental procedure as well as later during analysis of the data provided by the magnetometer.
This is a particularly important remark for everyone who investigates 'composite' specimens in which the signal in question constitute a small fraction of the total moment seen by the magnetometer. It can be either a (very) thin 'magnetic' layer deposited on the (typically) thick substrate, or a nano- or bio- object in a gelatin capsule or other container. It is a common wisdom, as well as the manufacturer suggestion, to measure in such cases, at least once, the sample 'carrier' separately and subtract this information from the measurements performed on the whole assembly.

In rare cases,  the specimen is supported by a substrate, which has a pure diamagnetic response, i.\,e., field and temperature {\it in}dependent) diamagnetism, the substrate's contribution can be determined from the measured data of the composite sample itself and subsequently being subtracted. However, this procedure is only adequate in cases where the substrate is devoid of paramagnetic impurities (usually glass or MgO are full of these). Further, they should not have a small band-gap (like Si or GaAs), since it gives rise to a temperature-dependent paramagnetic contribution stemming either from a van-Vleck type contribution due to the temperature dependence of the band-gap \cite{Ney:2006_JPCM} or from Pauli-like contributions stemming from temperature dependent carrier concentrations. Although these contributions are tiny for the substrate's overall magnetic response, after subtraction of the pure (i.\,e., field and temperature {\it in}dependent) diamagnetism, they can still be of the same order as the signal of the actual thin film or nano-object of interest. Typically, from the group of common substrate materials, only sapphire or SiC fulfill the above criteria; however, great care is still in order to avoid magnetic contamination, which can originate at any stage in-between growth of the single crystal and mounting the sample for the magnetometric study (see below).

Whichever way performed, this procedure usually leads to a high degree of "compensation", i.\,e. two large numbers (with limited accuracy) of approximately the same size are subtracted to yield the desired small magnetic response of the actual sample in question. Thus, a high absolute precision of the data is a must. The key point in measuring weak moments is establishing of the absolute credibility of the data provided by the MSMS system. Before doing so, we remark here that throughout the paper we disregard the obvious influence of the experimental noise because low as well as high frequency components can sufficiently be averaged out from $V(z)$ by multiple scanning and position averaging of the data. Consequently, for the sake of simplicity, we treat the data as noise-free and of a sufficiently low standard deviation.

\subsection{Magnetic moment coupling}
\label{sec:MgnMomCoupl}

\subsubsection{Sample size effects}
\label{sec:SampleSize}

\begin{table}[t]
\centering
\caption{Correction factors for MPMS XL SQUID magnetometers calculated for the most frequently met cuboid-shaped specimen sizes. $a$ and $b$ denote in-plane dimentions, $w$ is the thickness. The correct value for the moment of an investigated specimen is obtained when the MPMS reported value for the "longitudinal moment" is \emph{divided} by the relevant number from the table. }

\begin{ruledtabular}
\begin{tabular}{|c|c|c|c|c|c|}
\hline

\multicolumn{3}{|c|}{Sample size} & \multicolumn{3}{c|}{Coefficients} \\

\multicolumn{3}{|c|}{(mm)} & \multicolumn{3}{c|}{along the relevant side} \\ \hline

\multicolumn{3}{|c|}{  } & \multicolumn{2}{c|}{in-plane} & perpendicular \\ \hline

$a$ & $b$ & $w$ & $a$ & $b$ & $w$ \\ \hline

3.0 & 3.0 & 0.0 & 0.9938 & 0.9938 & 1.0113 \\ \hline

3.0 & 4.0 & 0.0 & 0.9860 & 0.9986 & 1.0156 \\ \hline

4.0 & 4.0 & 0.0 & 0.9891 & 0.9891 & 1.0219 \\ \hline

4.0 & 5.0 & 0.0 & 0.9946 & 0.9774 & 1.0280 \\ \hline

5.0 & 5.0 & 0.0 & 0.9831 & 0.9831 & 1.0341 \\ \hline

5.0 & 6.0 & 0.0 & 0.9902 & 0.9689 & 1.0416 \\ \hline

6.0 & 6.0 & 0.0 & 0.9758 & 0.9758 & 1.0491 \\ \hline

3.0 & 3.0 & 0.3 & 0.9939 & 0.9939 & 1.0122 \\ \hline

3.0 & 4.0 & 0.3 & 0.9985 & 0.9847 & 1.0170 \\ \hline

4.0 & 4.0 & 0.3 & 0.9892 & 0.9892 & 1.0217 \\ \hline

4.0 & 5.0 & 0.3 & 0.9951 & 0.9775 & 1.0278 \\ \hline

5.0 & 5.0 & 0.3 & 0.9832 & 0.9832 & 1.0339 \\ \hline

5.0 & 6.0 & 0.3 & 0.9902 & 0.9690 & 1.0414 \\ \hline

6.0 & 6.0 & 0.3 & 0.9759 & 0.9759 & 1.0489 \\ \hline

3.0 & 3.0 & 0.5 & 0.9940 & 0.9940 & 1.0120 \\ \hline

3.0 & 4.0 & 0.5 & 0.9986 & 0.9848 & 1.0167 \\ \hline

4.0 & 4.0 & 0.5 & 0.9893 & 0.9893 & 1.0214 \\ \hline

4.0 & 5.0 & 0.5 & 0.9952 & 0.9776 & 1.0279 \\ \hline

5.0 & 5.0 & 0.5 & 0.9833 & 0.9833 & 1.0336 \\ \hline

5.0 & 6.0 & 0.5 & 0.9904 & 0.9691 & 1.0411 \\ \hline

6.0 & 6.0 & 0.5 & 0.9760 & 0.9760 & 1.0486 \\ \hline

\end{tabular}
\end{ruledtabular}
\label{Table:CorrFact}
\end{table}

The point dipole-based approach to $\Upsilon(z)$ taken by Quantum Design (as well as other manufactures) has got profound implications for magnetometry as in the vast majority of measurements the specimens are up to 1 mm thick rectangulars of an area up to $5 \times 5$~mm$^2$, which are not necessarily ideally symmetric and perfectly centered in the gradiometer. All these factors modify the shape of $V(z)$ and by that influence the computed values of $m$. We start our analysis from size related effect and comment on the influence of the sample positioning in the next section.

The problem of non-zero specimen dimensions and their influence on the strength of the coupling with pick-up coils (this affects both amplitude and the shape of the $V(z)$) has been recognized already very early \cite{Zieba:1993_RSI,Miller:1996_RSI,Stamenov:2006_RSI} and the reader can find plenty of correction factors for various experimental configurations in these papers. However, these calculation are performed for objects of relatively high symmetry (spheres, squares, discs, etc.). For completeness, we list correction factors for cuboids of the most frequently used sample dimensions in Table~\ref{Table:CorrFact}. In our calculations these bodies are approximated by a set of point dipoles evenly occupying the whole volume of the specimens and a superposition of fluxes created by each individual element to determine the shape of $V(z)$. We also assume here that these bodies are uniformly magnetized in the direction of the axis of the symmetry of the gradiometer (the $z$-axis). Since in such conditions $V(z)$ scans are expected to be symmetrical with respect to the center of the gradiometer (position $z_0=0$ in Fig.~\ref{fig:SamplePos}) contrary to results presented in \cite{Stamenov:2006_RSI} both linear and iterative regression modes (the fitting fixed either at the center of the scan, or allowed to 'float' along $z$ to chase the "peak position", respectively) should give the same results (assuming ideally centered sample at $z_0$). The correction-factors which are smaller than unity for the "parallel" configuration originate from the vertical extent of these rather thin plates. Compared to an ideal point dipole centered within one of the windings of the gradiometer, a sample with a spacial extent along the gradiometer axis will induce a larger superconducting current of opposite sense in the other, more remote winding which leads effectively to 'flattening' down the resulting $V(z)$. A different effect is responsible for correction-factors larger than unity for the "perpendicular" configuration, when the sample extends more in the $xy$ plane than along the axis of the gradiometer. We know that the strength of the magnetic field produced by a current loop substantially increases when we move away from its center towards the rim. Thus, along the reciprocity principle, the coupling of a magnetic dipole with the sensing coil increases, resulting in a 'taller' $V(z)$. This case is exemplified in Fig.~\ref{fig:SamplePos}(b) where the test moment is placed 3~mm away from the axis of the gradiometer ($r/a \cong 0.33$). An apparent increase of $m$ of about 6.5\% will result from this particular case.

Although the numbers listed in Table~\ref{Table:CorrFact} fall to a range of small percents, and so they may be disregarded in rough test measurements, they are playing a decisive role in small moments determination, when a high degree of compensation is required.
To illustrate this effect let us consider an orientation dependent studies of a magnetically weak and submicrometer thin layer deposited on a macroscopic diamagnetic substrate. DMS such as nitrides or oxides doped with transition metals deposited on substrates are a good example of such samples. Let's further assume that we know the diamagnetic moment $m_{\mbox{\emph{dia}}}(H)$ of the substrate from a test measurement of a reference piece performed in only one orientation. Now, assuming $5 \times 5 \times 0.5$~mm$^3$ dimensions of the specimen, the difference between parallel and perpendicular configurations gives us about 5\% difference in SQUID response (see Table~\ref{Table:CorrFact} and \cite{Ney:2008_JMMM}, Fig.\ 4). If the magnetic response as a function of orientation is then established by a simple subtraction of this reference $m_{\mbox{\emph{dia}}}(H)$, an $H$-dependent $0.05 \cdot m_{\mbox{\emph{dia}}}(H)$ value will be artificially added (or subtracted) to the magnetic moments of the layer in question in the other configuration. Remarkably, since the saturated moments of these magnetically diluted layers are of the order of only few percent of the high field moments of their substrates, disregarding the effect of sample geometry may dwarf their real magnetic response.

\begin{figure}[th]
   \begin{center}
   \includegraphics[width=0.95\columnwidth]{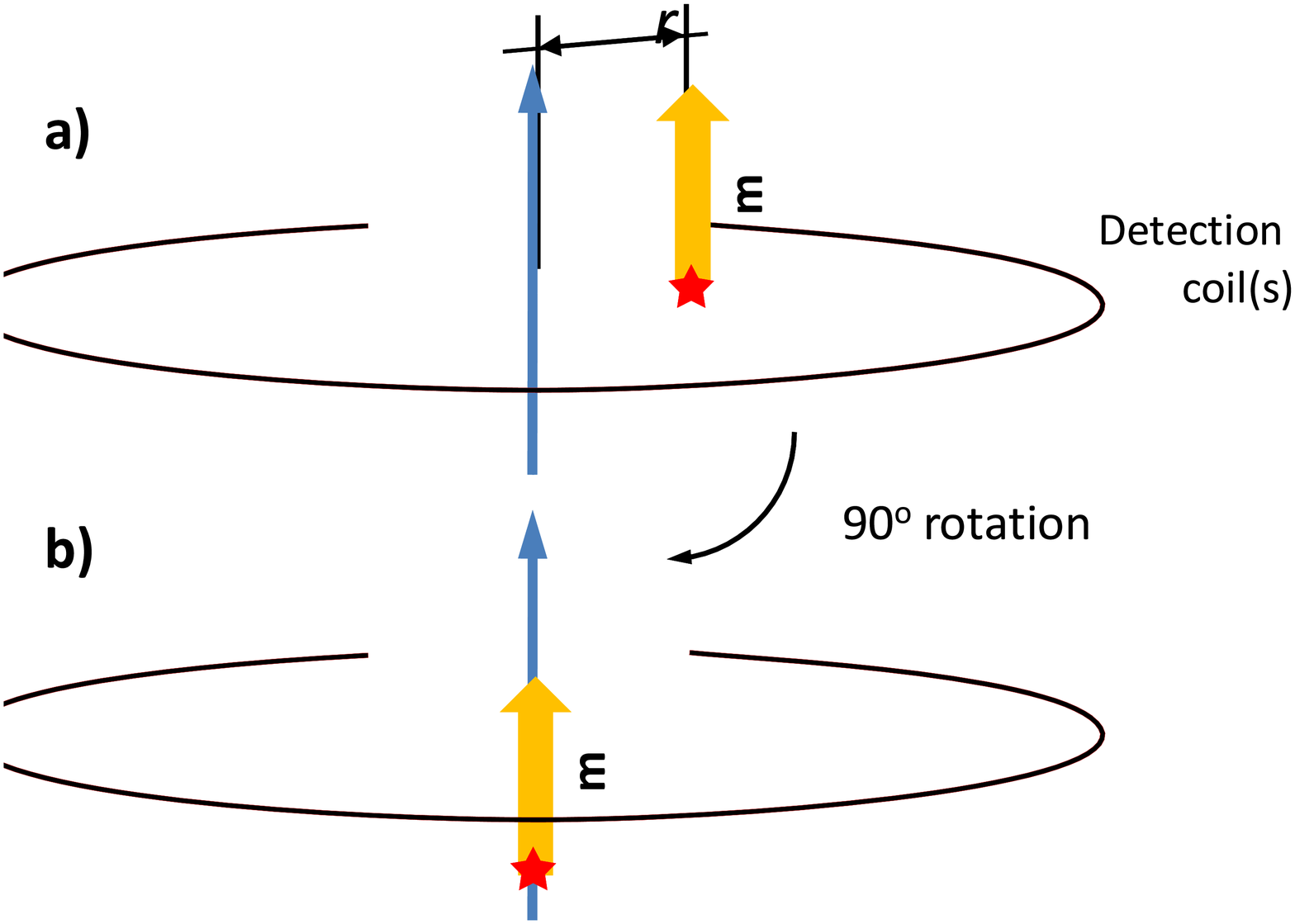}
   \end{center}
   \caption{(Color online) Two in plane orientations giving the same difference in saturation moments for the edge contaminant (the red star) as it would be between in-plane and perpendicular orientations. Importantly, the radial offset $r$ of the contaminant in (a) will be the same in the perpendicular case (not shown).}
   \label{fig:EdgeContaminant}
\end{figure}

Recently Garcia et al. \cite{Garcia:2009_JAP} suggested that a point ferromagnetic contaminant located at an edge of a millimeters sized sample which changes its $xy$-position upon mounting in different orientations is responsible for large differences in saturation values, $m_S$, seen in various orientation-dependent SQUID studies \cite{Carmeli:2003_JCP,Venkatesan:2004_N,Venkatesan:2004_PRL,Hernano:2006_PRB}. It was argued that rotating the sample to measure in-plane and perpendicular orientations changes the position of the hypothetical contamination particle with respect to the axis of the gradiometer. This should yield as much as 40\% difference in $m_S$ values if that particle is located on an edge of a $3 \times 3 \times 0.5$~mm$^3$ diamagnetic substrate, see Fig.~10 in \cite{Garcia:2009_JAP}. Following that scenario the edge contamination was far from the symmetry axis of the gradiometer during perpendicular measurement, while mounted for in-plane measurement it found its place on the axis, as sketched in Figs.~\ref{fig:SamplePos}(b) and (a), respectively. Although generally plausible, this effect can not produce a sufficiently large anisotropy in any MPMS system. The inner diameter of the sample chamber is about 9~mm, so the largest possible distance of a magnetic particle from the coils' axis is $r = 4.5$~mm (or $r/a \simeq 0.5$) for which the increase of the coupling is only 17\%. A 40\% effect would require an $ r/a \simeq 0.6$, as calculated by the same authors, see Fig.~9, which is a location already outside the sample chamber. We finally note that for the sample studied in \cite{Garcia:2009_JAP} $r/a$ is 0.15 for contamination at center of an edge or 0.22 for location at the corner, which gives at most 2\% difference for MPMS systems. Moreover, we find it highly fortunate to put this single edge contamination exactly at the magnetometer axis to see the maximum effect. In more realistic cases, as it is very hard to impose only a single contamination during handling, only smaller effects can be expected. The more evenly the contaminations are spread on the edges, the less difference in their $m_S$ will be seen between any two orientations. One should note, that the above effect is not only present in in-plane versus out-of-plane measurements. Also an in-plane rotation by 90$^{\circ}$ will create a similar effect, as it is exemplified in Fig.~\ref{fig:EdgeContaminant} but requires a less sophisticated sample mounting, i.~e.~additional contamination will be less likely. Interestingly, as it is elaborated in chapter~\ref{sec:HandlingArtefacts}, the experimental arrangement depicted in Fig.~\ref{fig:EdgeContaminant}(b) is very likely to produce a characteristic 'kink' in $m(H)$ and a up-down jumps of the sample position reported by the MPMS at $H$ where both substrate and contaminant signals compensate, see Fig.~\ref{SampleHandling}. We finally note that for typical sample sizes in an MPMS system the effect of edge contamination can account for maximum 6\% difference in $m_S$. Any larger anisotropy has to have a different origin.

To conclude this part we remark that for accurate magnetometry, the magnetometer response to the physical dimensions of the investigated bodies requires an accurate background compensation. In cases where the exact correction factors cannot be established numerically this requires reference measurements in basically all orientations and geometries in question.

\subsubsection{Positioning of the sample within the gradiometer}
\label{sec:Positioning}

As long as the sample is mounted on the sample holder in a way that it is placed exactly on the axis of the gradiometer, its exact position along the
gradiometer axis is irrelevant for an accurate determination of $m$ as it can always be centered either manually by the user or by a built-in automated routine (e.g., to compensate for the thermal expansion/contraction of the sample holder during temperature dependent studies). However, this is \emph{not} the case when the sample is mounted off the gradiometer axis. As already mentioned above, the strength of the $\mathbf{m}\leftrightarrow$~gradiometer coupling depends strongly on position of $\mathbf{m}$ with respect to center of the gradiometer coils, and so the correction factors listed in Table~\ref{Table:CorrFact} are only valid if the specimen is ideally centered on the gradiometer axis. This centering is usually assured by the sample holder, and commonly used drinking straws (provided with each MPMS system by Quantum Design) can assure adequate centering of the material, as sketched in Fig~\ref{fig:SampleInStraw}(a-b).

There are however at least three exceptions to this case. Firstly, drinking straws may not the best choice for accurate studies in nanomagnetism due to background signals, e.\,g., Ref.\ \cite{Bonanni:2007_PRB} and other designs may not offer adequate centering, see chapter~\ref{sec:SampleMounting}. Secondly, even in a drinking straw the sample may not be placed centrosymmetrically in cases the sample is mounted as sketched in Fig~\ref{fig:SampleInStraw}(c), or a curved straw will displace the sample from the center. Finally, we note that in nanomagnetism the investigated material is usually deposited on one side only of a bulky carrier, and so it occupies a different volume of the gradiometer than the substrate, and so a very precise absolute magnetometry would require calculation of separate correction factors both for the substrate and for the layer. As the latter is  very tedious, the best suggestion would be to investigate structures as similar as possible under very similar conditions, i.e. to use nearly identical and adequately tested sample holders and substrates. These allows to significantly reduce the number of control experiments with bare substrates and yet allows carrying out accurate magnetometry.

\subsubsection{Tilted moments (magnetic anisotropy)}
\label{sec:MgnAni}

So far, we were considering only the case of vertically (along the $z$-axis) magnetized samples. A ferromagnetic anisotropy present in the investigated nanostructure (or less favorably existing in the whole specimen), or a presence of a transverse component of the external magnetic field can rotate $\mathbf{m}$ away from $\mathbf{z}$. A transverse component of the magnetic field may appear in the superconducting magnet when it is near its remnant state, or when the system lacks a proper magnetic shielding so that the Earth's magnetic field, and/or a field produced by magnetized ferrous equipment  or reinforced concrete bars close-by, can leak into to sample chamber. As the latter cases are system- or environment-dependent, we leave them aside from more general considerations. On the other hand, the magnetic anisotropy itself is very frequently the actual object of interest so this case will be considered in greater detail in the following.

\begin{figure*}[th]
   \begin{center}
        \includegraphics[width=0.85\textwidth]{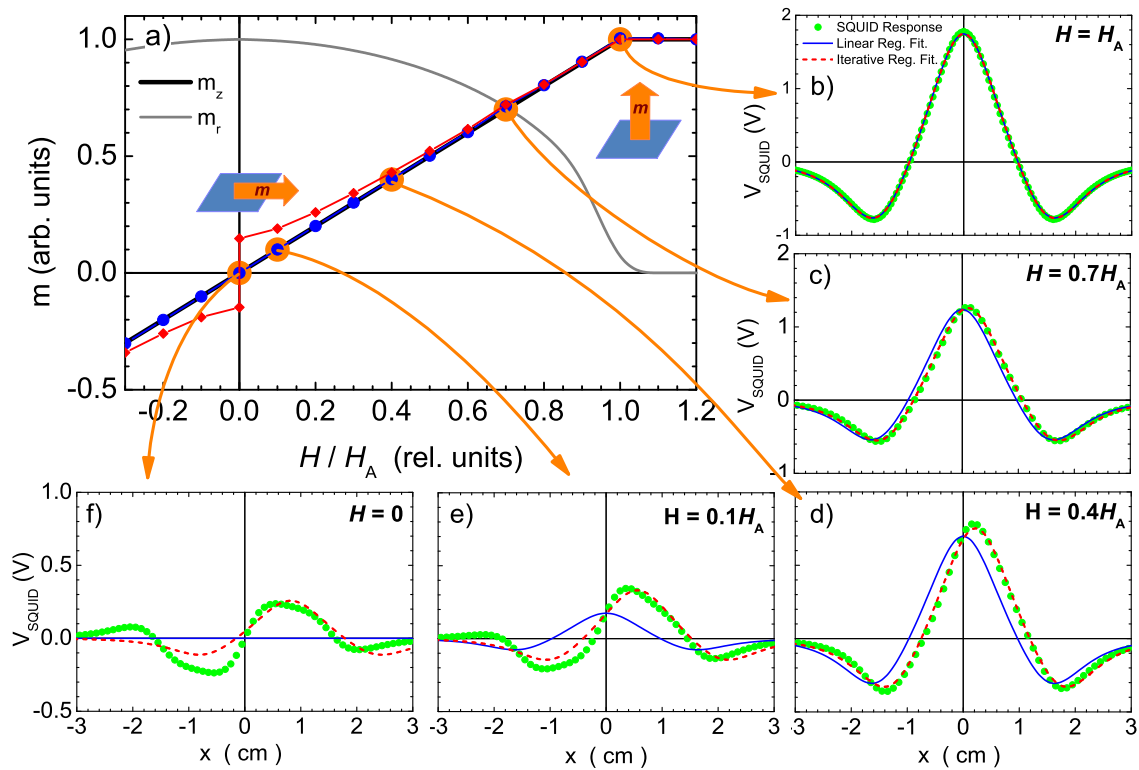}
   \end{center}
   \caption{(Color online) Magnetization reversal process along the hard axis in a thin ($d=$~0), macroscopically large ($5\times 5$ mm$^2$), and uniformly magnetized (a single domain state assumed) layer exhibiting very strong uniaxial magnetic anisotropy as seen by a SQUID magnetometer equipped with 2nd order gradiometer. Experimental configuration is depicted in the main panel (a). The thick black line represents the $z$ component while the grey line the radial component of \textbf{m} in this case. Both $m$ and $H$ are given in the relative units. Panels (b-f) depict the scans $V(z)$  (green points) calculated at five characteristic values of $H / H_A$. Then the ideal response function $\Upsilon(z)$ is fitted in both iterative (red dashed lines) and linear (solid blue lines) modes and the results are presented in the panel (a) as red diamonds and blue circles, respectively.}
   \label{fig:MgnAni}
\end{figure*}

\begin{figure}[h]
   \begin{center}
        \includegraphics[width=0.85\columnwidth]{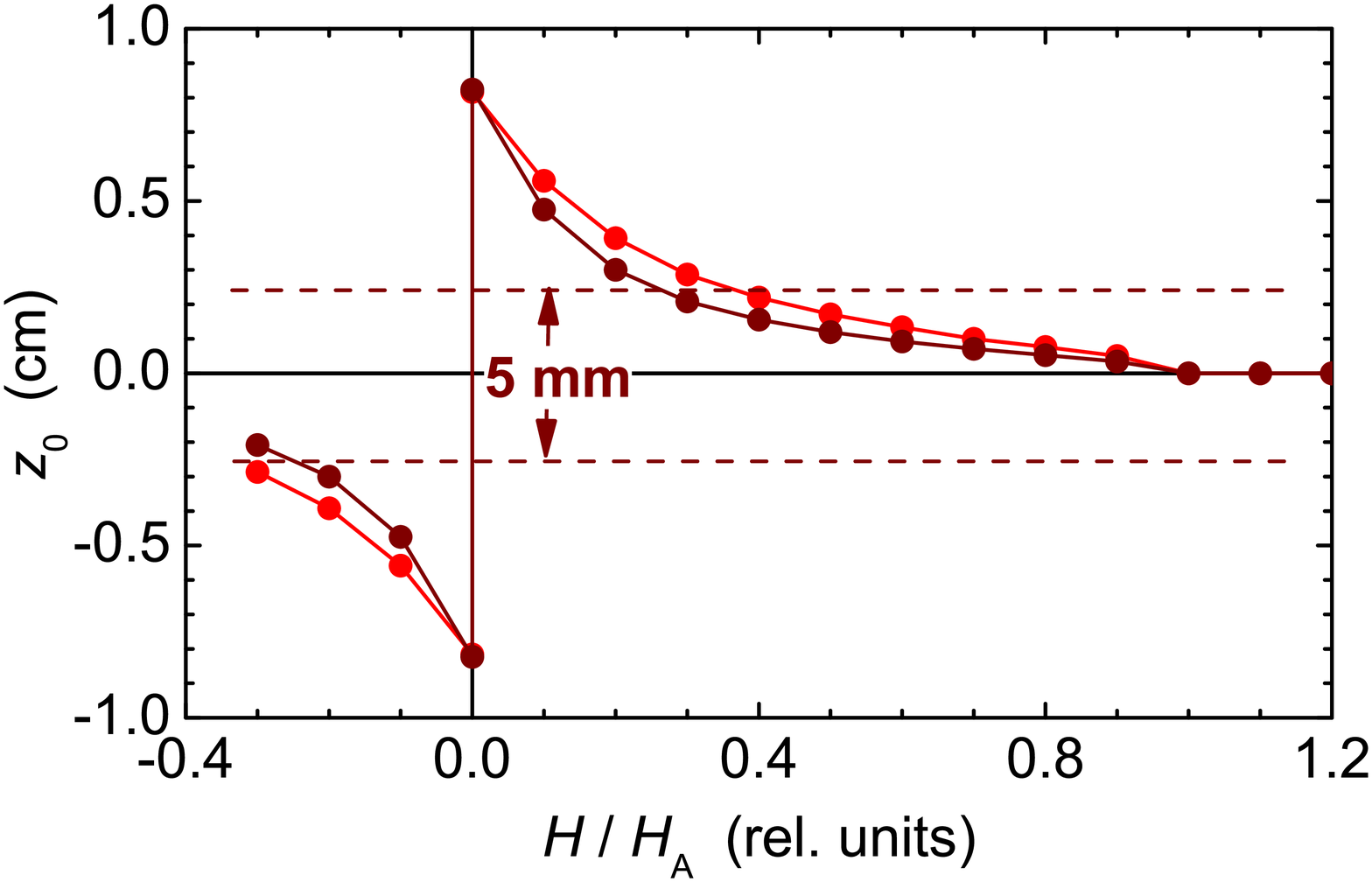}
   \end{center}
   \caption{(Color online) Red: offset of the scan or the "sample position" with respect to the center of the gradiometer calculated in the iterative regression mode for the perpendicular case considered in Fig.~\ref{fig:MgnAni}. Brown: results for the in-plane case are given. The two horizontal dashed lines mark the border of the physical extent of a $5 \times 5$ mm$^2$ sample.}
   \label{fig:z0UniAni}
\end{figure}

In the axially symmetric gradiometer pick-up coils every magnetic moment can be decomposed into its vertical, $\mathbf{m_z}$, and radial, $\mathbf{m_r}$, components. Again, from the reciprocity principle, we know that $\mathbf{m_r}$ will not couple to the SQUID as long as it is located anywhere on the symmetry axis of the pick-up coils ($r=0$). On the other hand, although $H_r$, the radial component of the magnetic field generated by a circular current loop, vanishes in the plane of this loop, $\mathbf{m_r}(r\neq 0)$ will be always sensed by the SQUID's gradiometer due to both the vertical movement of the sample (the principle of operation) and a vertical separation of the constituent coils of the gradiometer. This situation is exemplified in Fig.~\ref{fig:SamplePos}(c) indicating the general regularity that a presence of nonzero $\mathbf{m_r}$ anywhere outside the gradiometer axis of symmetry produces an odd $V(z)$. For a general case when both $\mathbf{m_z}$ and $\mathbf{m_r}$ are present at a certain volume inside the gradiometer, $V(z)$ acquires both even and odd components of a relative strength depending of the tilt angle of $\mathbf{m}$ and a degree of the magnetic mass distribution in the gradiometer coils plane (the $xy$ plane).

Let's illustrate this issue considering a magnetization reversal process along the hard axis in a thin ($d=$~0), macroscopically large ($5\times 5$ mm$^2$), and uniformly magnetized (a single domain state assumed) layer exhibiting very strong uniaxial magnetic anisotropy. (Ga,Mn)As, a member of DMS family, is a good example of such a material. It can exhibit either perpendicular magnetic anisotropy (PMA) with the easy axis oriented along normal to its plane, or in-plane magnetic anisotropy (IMA) with the normal being the hard axis, depending on the Mn content, hole concentration and temperature {\cite{Dietl:2001_PRB, Sawicki:2006_a}. As already argued before, stronger effects are expected when the sample is mounted in perpendicular configuration that is when the layer is parallel to the coils plane. We consider IMA case, so without extrnal magnetic field the magnetic moment of the layer is oriented in the sample plane. Then $m_z=|\mathbf{m_z}|=0$, $m_r = |\mathbf{m_r}| = m$, and the external field is needed to tilt and finally align $\mathbf{m}$ with the $z$~axis when $H$ becomes stronger than the anisotropy field $H_A$ of this layer. For the uniaxial magnetic anisotropy $m_z = m$ for $|H| > H_A$, and $m_z = m\cdot H / H_A$ for -$H_A < H < H_A$, and a response linear in $H$ is expected for $m_z$.
Fig.~\ref{fig:MgnAni}(a) depicts the experimental configuration and shows the expected experimental response (bold black line). In panels (b)-(f) of the same figure green points depict calculated gradiometer responses in the presence of nonzero $m_z$ and $m_r$ at 5 different values $H / H_A$.
Then we fit $\Upsilon(z)$ in both linear (solid blue lines) and iterative (dashed red lines) regression modes and plot the two obtained sets of $m$ in the panel (a) as line-connected solid points of the same colors.

Although at high fields the results of both methods coincide, one immediately notices a significant difference in the $m$ values yielded by these two modes of fitting at low magnetic fields. Counterintuitive, the iterative mode which allows for reasonable good fits of $V(z)$ in the whole range of magnetic fields yields progressively worse numbers as the $m_r / m_z$ ratio for this macroscopically large specimen increases. The red set of points reproduces neither the $m_r$- nor the $m_z$- dependence on $H$ at weak fields. Actually, the magnitude of $m$ reported by the iterative regression mode for large $m_r / m_z$ ratios depends strongly on the range of $z_0$ values allowed for the fitting routine. At the same time, the linear mode, which does not fit the computed $V(z)$ well at low magnetic fields, especially at $H=0$, yields values of $m$ which perfectly reproduce the expected values of $m_z(H)$ in whole range of $H$. In other words, the $V(z)$ does not reflect solely the magnetic response of the specimen but also comprises artifacts stemming from the finite sample size and the gradiometric detection system.

How can we know now that moment tilting is in force? If someone decides to use the iterative regression mode for $m(H)$ loops at different temperatures then the signature of the maoment-tilting is included in the information on the fitted "best" sample position. This can be observed during the measurement or conveniently been read out afterwards from  "Long Offset (cm)", the 7th column of the *.rso.dat file. Fig.~\ref{fig:z0UniAni} collects the "best" sample positions of the iterative regression mode corresponding to the example presented in Fig.~\ref{fig:MgnAni}. The characteristic "divergence" around $H=0$ and a sudden change of sign indicate the presence of a substantial odd component in $V(z)$. For comparison, Fig.~\ref{fig:z0UniAni} shows also the "best" center position for a similar sample investigated in-plane. Then largest $m_r / m_z$ ratio will occur near $H=0$ when the system exhibits PMA or when a system with a strong in-plane uniaxial anisotropy is measured along its hard axis, or oriented at any direction at $H$ near the coercive field when the magnetization reversal occurs through coherent rotation of the whole \textbf{m}. Please note two features in this figure: (i) although the artificial off-centering is smaller for the parallel case, the divergence is equally strong in both cases; (ii) the apparent shift of the sample position during the $m(H)$ loop exceeds the real sample dimensions. These features are observed despite the fact that the sample cannot change its position during the measurement at constant temperature. The only remedy in this case is to recalculate the raw data having the sample position fixed for the whole $m(H)$ at the \emph{real} sample position, established, for example, from high field data. The only way to avoid recalculation is to set the system to work in the linear regression mode, but this require proper centering the sample before each field loop, so to compensate for the temperature expansion of the sample holder. In general, we may not retrieve 100\% accurate values, but at least the odd component will not severely mar the outcome.  Note, that the discussed here feature is very similar in appearance to the edge-contamination presented in Fig.~\ref{SampleHandling}(c).

\subsection{The magnetic field}
\label{sec:MgnField}

The ability to precisely control the values of the external magnetic field acting on the studied specimen is one of the most important features of any useful magnetometer. The MPMS XL system offers magnetic fields up to 50 or 70~kOe which are generated in a high-homogeneity superconducting coil which is energized by a high current power supply controlled by the MPMS software. However, there are some issues related with the generation of the magnetic field that influence the fidelity of the final outcome of the magnetometer. The most severe cases may drive the inadvertent scientist to false or even unphysical conclusions. These issues are related to both the magnet power supply and to the magnet itself, and will be discussed in that order. We also exemplify an effect of a substantial reduction of the MPMS accuracy seen at magnetic fields stronger than about 25~kOe. As we cannot unambiguously identify its origin, we will discuss it at the end of this section.

\subsubsection{Magnet power supply}
\label{sec:MgnPowerSupply}

The contemporary power supplies offer a very high degree of linearity of their voltage-to-current (input-to-output) conversion. The current for the magnet of the MPMS magnetometer is provided by a unipolar power supply (Kepco). The very important point to be noticed by any user of the system is that the actual magnetic field \emph{is not} measured directly by a field sensor. Instead, the field values are determined by the voltage-drop over a shunt resistor which is amplified by a factor of 21 before it is compared to the set voltage of the control board to select the desired magnetic field. The difference of the two voltages is fed into the Kepco as control voltage. This magnetic field control has to be well-calibrated on a regular base and procedures are available from Quantum Design both for the voltage amplifier and the zero-current offset of the Kepco.

To cover the wide field range of the MPMS there are two shunt resistors with respective amplifiers between which the system is switching when is moving from high resolution to low resolution charging mode. The specific field of switching is related to the field/current magnet calibration factor (unless forced by the user-defined sequence), but always within region near few kOe. It is factory calibrated to better than +/- 0.5\% but the exceptional precision of the SQUID and relative charging accuracy of the Kepco enable the step to be clearly seen. In particular the switch is marked in Figs.~\ref{fig:TrappedField} and \ref{fig:CleanHtr} by faint dashed lines (this MPMS switches just above 5~kOe). Ideally, the adequate calibration should remove the step, practically, one can only minimize it by applying slightly different correction factors at both field ranges. The other, perhaps more practical, solution is to treat the slightly wrongly reported field values as an additional contribution to a remnant field in the magnet and correct both effects simultaneously, as described in the next section.

It should be further noted that the magnetic field control comprises a self-correcting routine which is intended to correct small discrepancies between the actual current in the magnet and the current set by the power supply which can stem from moderate flux relaxation. In cases of field losses of the magnet, malfunction of the persistent-switch, or severe flux creep in the magnet, this self-correcting mechanism may lead to increasingly larger correcting factors resulting in very inaccurate nominal field values compared to the real ones. This can result in large residual hysteresis or even inverted artificial hysteresis. To remove these issues, the 1822-unit (DSP board) of the MPMS has to be switched off for at least 10 sec (the message: "`power-on reset occurred"' has to be flagged). This procedure which resets the magnetic field correction routine is also necessary before the zero-current offset of the Kepco is corrected.

\subsubsection{Superconducting coil}
\label{sec:SuperCoil}

As stated above, there is no absolute field sensor located anywhere in the magnetometer, so the magnetic field value reported to the user is based only on the current set by the power supply. Because a superconducting solenoid is used to generate the magnetic field, the user has to take into account the existence of pinned magnetic flux lines and flux movement within the magnet. In general, this makes it difficult to precisely determine the actual magnetic field at the location of the sample. This is particularly true at weak magnetic fields. There are two phenomena that cause a discrepancy of the MPMS nominal field values (which is laid down in the data files) from those really experienced by the sample. Namely they are superconducting magnet remanence and flux creep and escape \cite{QD1070-207}. In the following we will discuss the main implications of these effects for the studies of small-signal samples and suggest methods of mitigating these artifacts.

The remnant or trapped field in the superconducting magnet, $H_{tr}$, results from the penetration of the material of the magnet by quantized magnetic flux. This flux gets pinned at defects in the material and when the magnet is discharged some of these flux lines remain pinned, giving a weak magnetic field at the sample location. Importantly, the spacial distribution of $H_{tr}$ is very complicated (see for example \cite{QD1014-208A}, where apart of a characterization of this problem in MPMS systems, some methods of achieving the lowest possible remnant fields are given). As in the RSO mode of measurements sample travels only in a very central part of the magnet bore, we can treat this field as an additional homogeneous field acting on the sample, but with very important characteristics: it is directed \emph{opposite} to the last experienced strong field by the magnet and assumes the largest values around zero magnetic field \cite{QD1070-207}. This results in a well known "negative" hysteresis, an apparent inverted (clockwise) magnetic field loop seen in soft ferromagnets when the results are plotted against $H_J$, the field \emph{calculated} from the current sent to the magnet and reported in the MPMS system's data files. This fictitious effect occurs only when this nominal $H_J$ is used instead of the real field $H_R$ taken from a field sensor located in the close vicinity of the sample (which does not exist in the MPMS). Since in this chapter we consider the effects caused by the difference between $H_J$ and $H_R$, we will strictly stick to this notation here. We abandon this distinction in the rest of the text and use the symbol $H$ which actually indicates $H_J$ (values given in the .dat files) unless it is clearly stated otherwise.

\begin{figure}[th]
   \begin{center}
        \includegraphics[width=0.9\columnwidth]{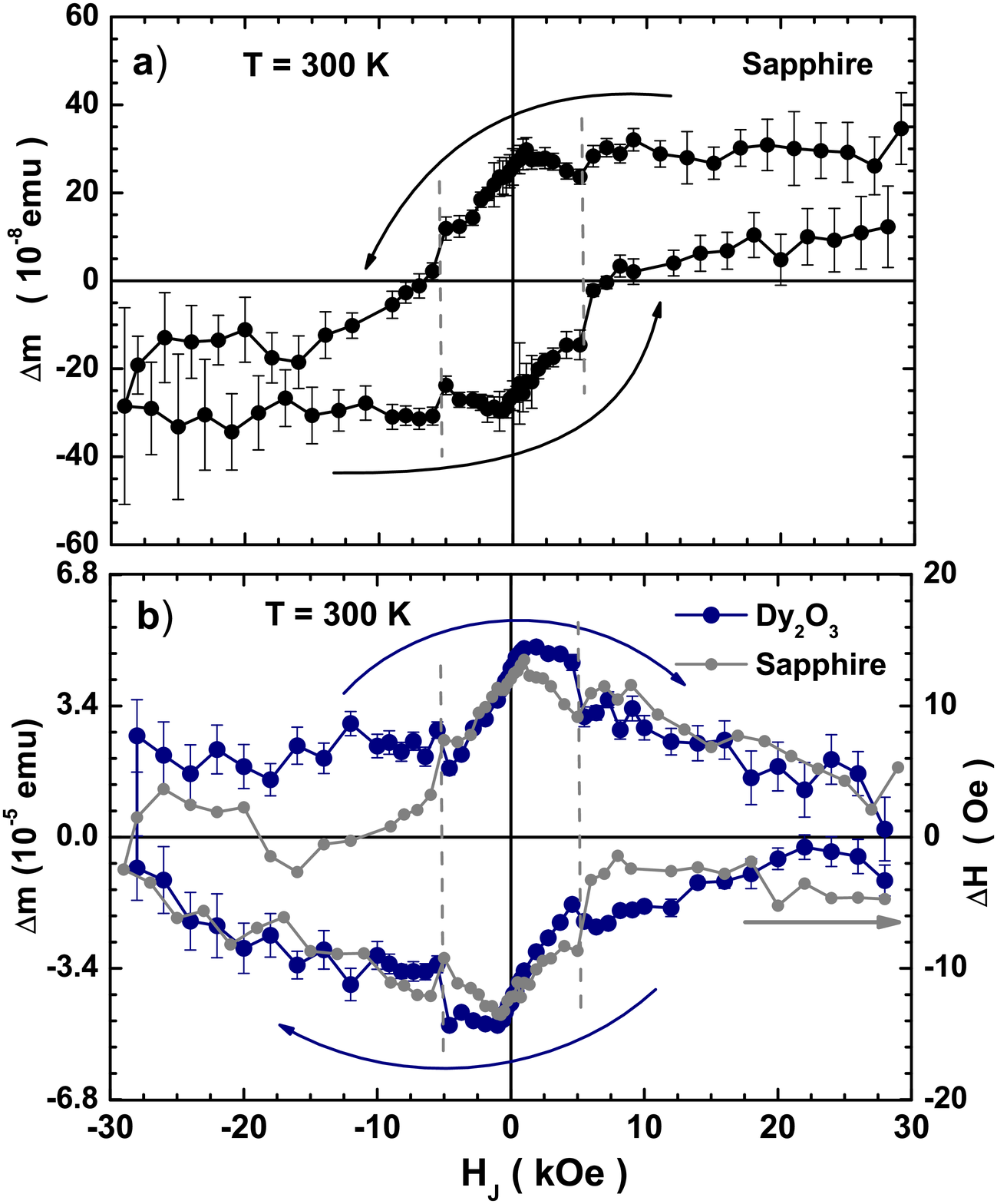}
   \end{center}
   \caption{(Color online) Left axes: nonlinear component with respect to $H_J$, the magnetic field reported in MPMS *.dat files which is calculated upon the current sent to the magnet, as a function of $H_J$ for (a) 60~mg diamagnetic sapphire substrate and (b) paramagnetic salt Dy$_2$O$_3$. The degree of compensation of original $m(H_J)$ exceeds 99.9\% above 10~kOe in both cases. Please note the opposite chirality of the two reported $\Delta m(H_J)$. The right axis in (b) denotes calculated values of the trapped field in the magnet, see the text. The dashed grey lines indicate the switching field when the power supply is switching from high resolution to low resolution charging mode.}
   \label{fig:TrappedField}
\end{figure}

The very existence of $H_{tr}$ (resulting in a discrepancy between $H_J$ and H$_R$) and the use of $H_J$ results in an apparent ferromagnetic-like response seen in samples which the magnetic response is dominated by a diamagnetic contribution. As the example of such an effect we discuss a room temperature hysteresis loop $m(H_J)$ collected for a sapphire substrate. It is important to note that there may exist a 'genuine' FM component in sapphire \cite{Salzer:2007_JMMM} and in particular in a non-properly handled sapphire \cite{Ney:2008_JMMM}. We discuss this issue further in this review (chapter \ref{sec:HandlingArtefacts}). Here we limit ourselves to magnetically clean material, which has been handled in the best possible way to avoid any magnetic contamination. The actual specimen for this test consists of two $5 \times 5 \times 0.3$ mm$^2$ pieces of total mass of 60 mg. The direct measurement gives fairly linear in $H_J$ values of the order of $2 \times 10^{-5}$ emu/kOe, so in order to check whether a ferromagnetic-like contribution is present we compensate the MPMS data by a linear, $\alpha H_J$ term, with $\alpha$ manually adjusted to flatten out the resulting dependency in a widest possible range of $H_J$. Therefore $\alpha$ can be regarded as a specific susceptibility of this specimen, $\chi_S$. The resulting $\Delta m(H_J) = m(H_J) -\alpha H_J$ is presented in Fig.~\ref{fig:TrappedField}(a). Although resembling a ferromagnetic response, such a shape is caused by the presence of the trapped flux $H_{tr}$ in the superconducting magnet. Other evidence for the dependence of this type of residual hysteresis in net-diamagnetic samples on the history of the magnet has been provided before \cite{Ney:2008_JMMM}. Here it will be demonstrated that identical residual hysteresis with opposite chirality can be recorded when paramagnetic samples are used.

To substantiate the origin of the residual hysteresis in sapphire and to determine the size of the residual field of the superconducting magnet we turn to the Dy$_2$O$_3$ sample (a paramagnetic salt) provided by Quantum Design for every MPMS system equipped with AC option\footnote{the Dy2O3 sample is used for the AC phase calibration. It is an electrical insulator with a very large paramagnetic susceptibility and in the frequency range from 0.001 to 1000 Hz its moment is in phase with the applied field ($\chi$-bis is zero). It is expected to provide highly linear response to the applied magnetic field, at least at elevated temperatures. Alternatively one can also use the Pd test sample provided by Quantum Design for samples without the AC option.}. We measure this test sample with the identical experimental procedure (i.\,e., identical history of the magnet) as the sapphire substrate and compensate for the linear term, $\beta H_J$, this time adjusting $\beta$ to obtain $\Delta m(H_J)$ symmetrical with respect to $Y$ axis at high fields \cite{QD1070-207}, see Fig.~\ref{fig:TrappedField}(b) dark blue points. Importantly, this time we note the clockwise character of $\Delta m(H_J)$ loop (an inverted hysteresis), the opposite one to that of the sapphire. Using the well-established value of $\beta$ we can calculate the \emph{real} magnetic field acting upon the sample $H_R = m(H_J) / \beta$ and obtain the "field error" $H_{Tr} = \Delta H = H_R - H_J$. The values $\Delta H$ are indicated on the right-axis in Fig.~\ref{fig:TrappedField}(b). To cross-check the validity of our approach, we repeat the same procedure to calculate $\Delta H$ for the sapphire substrate taking this time slightly altered value $\alpha$ as $\chi_S$. We add the results to Fig.~\ref{fig:TrappedField}(b) (marked in grey) and the full equivalence of both sets of $\Delta H$ confirms the common origin of the observed $\Delta m$: the extra field acting on the specimen, the $H_{tr}$ field. The different chirality of both $\Delta m(H)$ loops can now be readily understood, as $H_{tr}$ is the same positive or negative field for the sweeps up or down, respectively, acting on substances with an opposite sign of the susceptibility.

On the other hand it is very instructive to note that we needed a marginal change of $\alpha$ (less than 0.05\%), from $2.1465 \times 10^{-8}$~emu/Oe used to calculate $\Delta m$ in Fig.~\ref{fig:TrappedField}(a) to $2.1457 \times 10^{-8}$~emu/Oe used to asses $\Delta H$ in Fig.~\ref{fig:TrappedField}(b) to effectively remove the deceiving ferromagnetic-like appearance of $\Delta m$ presented in Fig.~\ref{fig:TrappedField}(a). It shows how easy it is to falsely interpret the data. It is therefore instructive to see what happens when one disregards $H_{tr}$ and considers such $\Delta m$ from Fig.~\ref{fig:TrappedField}(a) as a genuine ferromagnetic response. In this case the 'saturation' $m_S$ (or a 'remanence', if one considers the $H=0$ case) is $m \simeq 3 \times 10^{-7}$~emu. Let's imagine now that this is not a pure test-substrate but a typical sample consisting of a thin paramagnetic DMS layer deposited on a similar sapphire substrate. A recent investigation of $x < 1$\% (Ga,Mn)N is a good example of such a study, see \cite{Stefanowicz:2010_PRBa}. Taking a typical thickness of the DMS layer of $\sim 200$~nm one easily notices that $S$, the value of spin per each Mn is proportional to 1/$x$(\%). Therefore, calculating $S$ per Mn ion from the artificial $m_S$ of sapphire, $S$ exceeds the maximum expected value for Mn ($S = 5/2$) already at room temperature in such layers with $x < 0.4$\% even if the Mn is completely unpolarized. Remarkably, the effect will be stronger for thicker substrates, thinner layers and in nominally stronger magnets ($H_{tr}$ approaches 30~Oe for the 70~kOe MPMS).

On the contrary, a presence of a strong paramagnetic component in the whole signal of the investigated specimen, frequently met in DMS layers at low temperatures, will reduce either absolute values of the overall $m$ at low magnetic fields or a coercive force (if present). It also leads to 'negative' hysteresis loops if the true coercive force is in the few-Oersted range. More interestingly, the $H_{tr}$-induced effects will be strongly temperature dependent, and so they may create a number unusual and 'interesting' findings.
In any case, depending on the history of the field in the magnet, an $H_J$-based low field evaluation of any magnetic signal will be at least inaccurate, if not completely wrong.

Since the flux pinning is a property related to the type II superconductivity of the NbTi, the basic material from which the superconducting wire of the magnet is made of, the only way of minimizing $H_{tr}$ is by warming the magnet above 9~K, the superconducting critical temperature of NbTi. As in general this is highly impractical idea, there is a very useful option available, called "magnet reset", which, upon our experience in weak signal magnetism, we regard as a kind of "must have" for everyone interested in sensitive and precise magnetometry, at weak magnetic fields in particular. This option features one layer of heater wire wound along the entire length of the magnet and radially about half way through the magnet. A heat pulse through this heater will locally turn some windings into the normal (resistive) state which will then dissipate more energy (heat) to drive some neighboring windings to the normal state, and so on. Finally the whole superconducting material is driven into the normal state, and so the all pinned flux is expelled from the body of the magnet on an expense of a little amount of liquid helium in otherwise helium full dewar. It should be noted that this option is only useful in cases when a small magnetic field is still in the magnet. At nominally zero magnetic field the use of the reset option may not heat the entire magnet and---opposite to what is intended---$H_{tr}$ may increase or even reverse sign, see \cite{Ney:2008_JMMM}, Fig.\ 6.

Unfortunately, the magnet reset can purify the magnet exclusively for operation around $H=0$. It is of no help during typical $m(H)$ measurements, particularly those requiring the full extent of the magnetic fields. In such a case, we see no other method to eradicate $H_{tr}$ from post measurements analysis then by numerical replacement $H_J$ in data sets by $H_R$ using $H_{tr}$. To make it possible one has to establish the $H_{Tr}(H_J)$ dependence for the specific system and the field (and eventually temperature) history of the magnet.

\begin{figure}[t]
   \begin{center}
        \includegraphics[width=0.9\columnwidth]{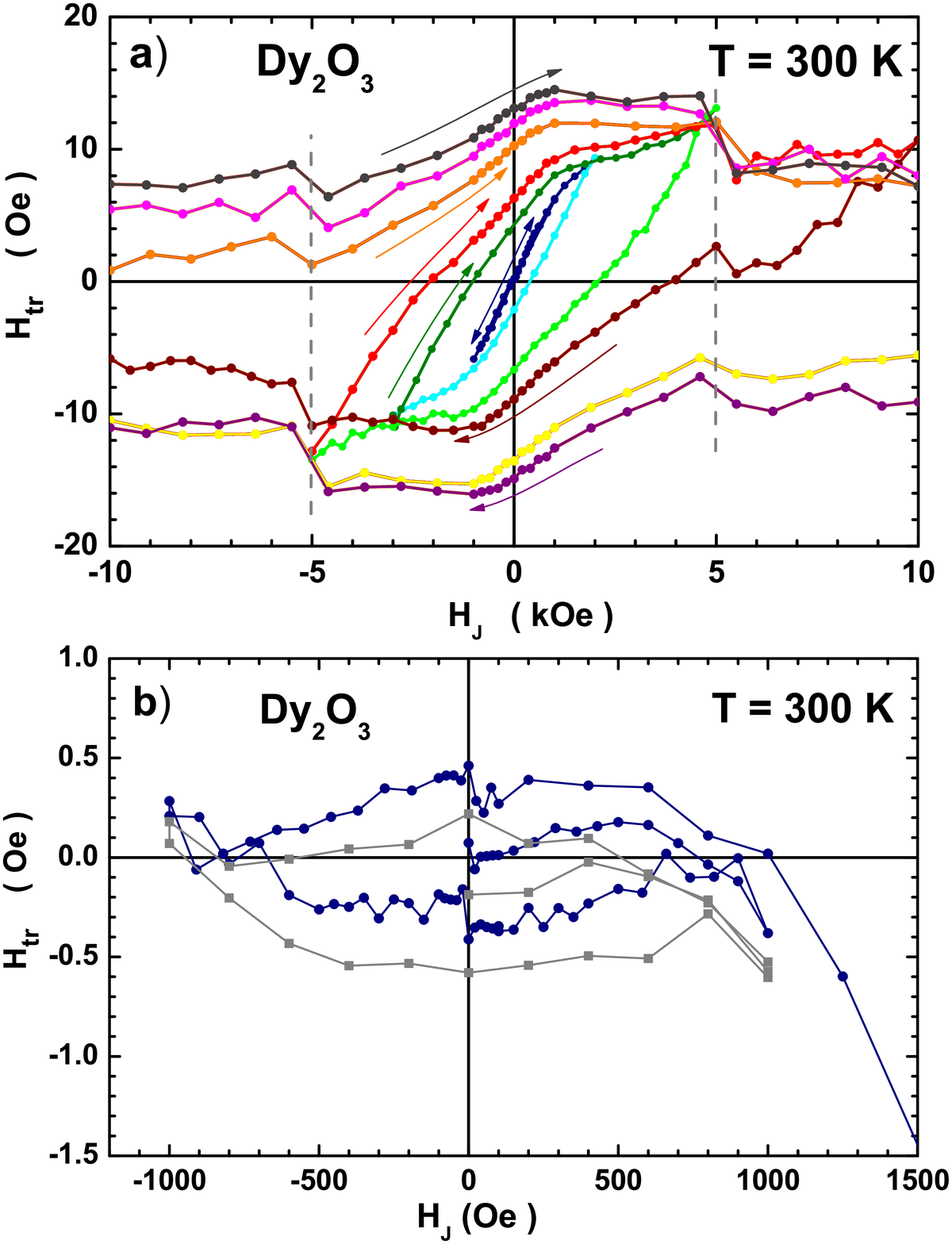}
   \end{center}
   \caption{(Color online) (a) A development of the remnant (trapped) field in an MPMS 5T magnet studied through sequential field sweeping to progressively stronger magnetic fields showed in $\pm 10$~kOe window. The initial part, $\pm 1$~kOe and to 2~kOe (dark blue) was measured after nulling the field by using with a self written degaussing sequence followed by the magnet reset. Then the field was swept to -3~kOe (light blue), 5~kOe (dark green), -5~kOe (green), 10~kOe (red), -20~kOe (brown). The remaining colors indicate field sweeps from: -20~kOe (orange), -20~kOe (orange), 32~kOe (yellow), -32~kOe (magenta), 50~kOe (purple), and  -50~kOe (grey). The remnant field was  calculated upon magnetic moment data of paramagnetic Dy$_2$O$_3$ described in the text. Arrows of the corresponding colors indicate the field sweep directions.  The dashed grey lines indicate the switching field when the power supply is moving from high resolution to low resolution charging mode. (b) Dark blue: the lowest field part of the data from (a). The grey marked points were collected after the whole sequence from (a) followed by magnet reset option. Note a small negative field, $H_R \simeq -0.2$~Oe, still present in the magnet. The data for this panel were compensated using slightly different factor to level them off at $\pm 1$~kOe range. }
   \label{fig:CleanHtr}
\end{figure}

The bottom-line is that there is no universal 'table' of $H_{Tr}(H)$ as $H_{tr}$ depends on the most recent magnet history. This is illustrated in Fig.~\ref{fig:CleanHtr}. We depict there $H_{tr}$ in $\pm 10$~kOe window established during a multiple field looping up to $\pm 50$~kOe through progressively stronger magnetic fields in the MPMS XL 5~T in Warsaw, Fig.~\ref{fig:CleanHtr}(a) (data presented in Figs.~\ref{fig:TrappedField}(b) and \ref{fig:HiFiRes}(a) are a part of this large data set). We started having the magnet purified with a self written degaussing sequence followed by the magnet reset. What can be immediately seen is a change of the initial 'slope' of the data in the field range to  $\pm 1$~kOe (dark blue) compared to the rest of the field sweeps. Importantly the slope is preserved as the field is limited to this range and then it starts to gradually change during the following field sweeps to 2, -3, 5, ..., and so to $\pm 50$~kOe. Importantly, if the same field loop is performed twice (not shown here), the obtained $H_{tr}$ values are very much the same. So, if a particular type of measurement is performed frequently, its pays off to established the specific set of $H_{Tr}(H_J)$ and use it to correct $H_J$ with a great confidence.
It is worth noting that this method also automatically compensates for a slight change of $H_J$ at $\pm 5$~kOe. A small jump in the data is seen in Figs.~\ref{fig:TrappedField} and \ref{fig:CleanHtr} and the relevant switching fields are marked by vertical grey dashed line. These jumps occur when the power supply is switching from high resolution to low resolution charging mode, see chapter~\ref{sec:MgnPowerSupply}.

A necessity of performing the degaussing procedure to establish a true zero field in the magnet is exemplified in Fig.~\ref{fig:CleanHtr}(b) where grey points mark $H_{tr}$ measured after performing the magnet reset alone after the described previously field sweeps to 50~kOe. For this panel we compensate the data by slightly different value of $\beta$ (increased by 0.6\%) to blow up a change of $H_{tr}$ during field sweeping in weak field range. We note that the 'grey' $H_{tr}(H_J)$ values are shifted down with respect to the initial ones, and the overall spread of the points (marginally weak hysteresis) covers the range of less than 0.5~Oe at fields below 1~kOe. This indicates that the absolute field reproducibility at these weak magnetic fields can be maintained better than 0.05\%.

\subsubsection{Flux creep and escape}
\label{sec:FluxCreep}

Even in the persistent mode of the magnet, some of the pinned magnetic flux lines in the superconducting wire will redistribute with time and some will leave the wire material. These that enter the bore of the magnet must be represented by an extra nonvanishing current in the magnet, so the magnetic field will not be stable due to the very nature of the type II superconducting material of the magnet (for an extensive description see \cite{QD1070-207}). This property has got a profound implication in low field magnetometry, e.g. when one wants to establish $\chi(T)$ of a paramagnetic nanostructure at high temperatures (see \cite{Bonanni:2010_arXiv}). Therefore a subtraction of $m(T)$ of the substrate measured at the same $H$ (normalized by the corresponding sample weight) must be made with the best possible accuracy. The time instability and history dependence of the magnetic field in the magnet renders this task impossible if one does not take the flux creep into account. The exact value of $H_R$ is history dependent and if one sets a weak filed $H = H_J$ the obtained $H_R$ may be up to $\pm 15$~Oe off the target (see Fig.~\ref{fig:CleanHtr}(a)). One can try to live with this by sticking to exactly the same experimental routine, assuring repeatable $H_R$, but the flux creep and escape cause the whole $m(T)$ to be performed at a \emph{continuously varying} magnetic field, as is indicated in Fig.~\ref{fig:FieldStab}, part (a). There the field was brought down to 1~kOe from 10~kOe in no overshoot mode resulting in a field drift on the magnet over hours. In this particular case, the magnet reset option was of great help (as indicated in parts (b) and (c) of the same measurement) by purging the magnet from the trapped flux directly before switching the 1~kOe field on. Thus it allows not only for setting a perfectly repeatable field values but also it sets conditions for an order of magnitude smaller field drift, independently of the most recent field history in the magnet.

\begin{figure}[th]
   \begin{center}
        \includegraphics[width=0.9\columnwidth]{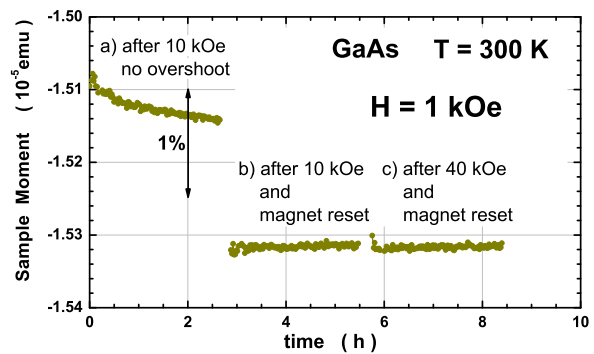}
   \end{center}
   \caption{(Color online) Field drift in 5~T magnet of MPMS system. The magnetic response at $H_J = 1$~kOe was recorded each time for approx. 3 hours for a piece of GaAs substrate after: (a) the field was brought down directly (in the no overshoot mode) from 10~kOe, (b) after the initial 10~kOe the magnet was firstly purified using the magnet reset option and then the 1~kOe was set, and (c) the same as (b), but with the starting field of 40~kOe. The negative $H_{tr}$ of the order of -15~Oe present in (a) is responsible for the vertical offset between (a) and (b-c).}
   \label{fig:FieldStab}
\end{figure}

\subsubsection{Limits of high field resolution}
\label{sec:Resolution}

As already pointed out, accurate studies in nanomagnetism require a great deal of compensation to be involved and obviously, the accuracy of a difference of two comparable numbers directly depends on the absolute precision of both the minuend and the subtrahend. As the system establishes $m$ on the account of a numerical comparison of $V(z)$ to $\Upsilon(z)$ \footnote{ There are three commonly used procedures for data reduction: a \emph{full scan} algorithm, a \emph{maximum slope} method, and a \emph{least-squares regression} see \cite{QD1014-214} and the discussion in ref \cite{Stamenov:2006_RSI}. We find that the third one is the most suitable in a reliable minute moments magnetometry, so this is our method of choice to establish $m$ and all the results presented throughout the text are obtained by least-squares regression fit of $\Upsilon(z)$ to $(V(z)$.}
then any deviation of the functional form of $V(z)$ from that of $\Upsilon(z)$ will yield somewhat distorted value of $m$. We routinely detect such distortions in all the scans obtained at magnetic field stronger than about 20~kOe.

\begin{figure}[th]
   \begin{center}
        \includegraphics[width=1\columnwidth]{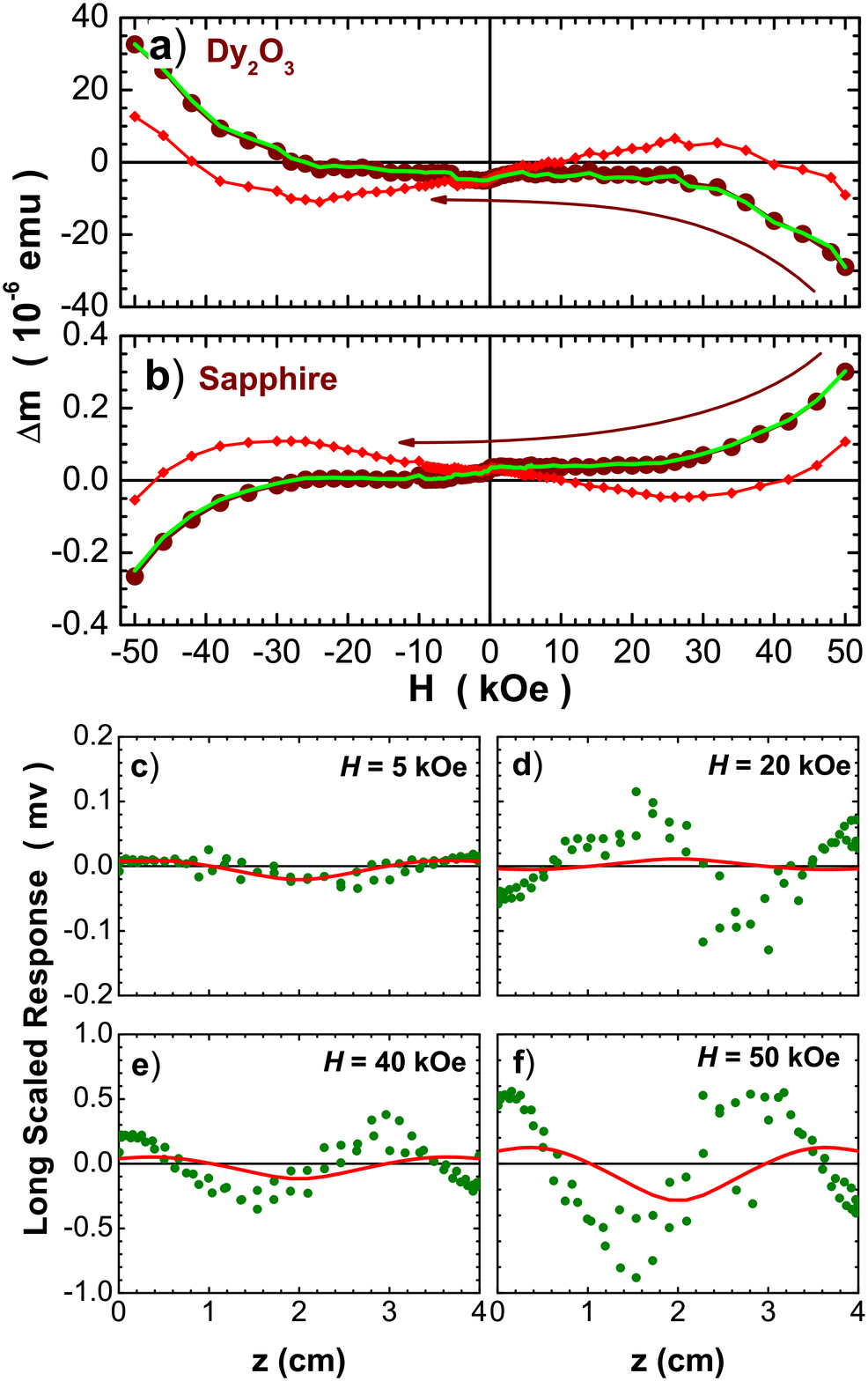}
   \end{center}
   \caption{(Color online) (a) and (b): (brown bullets) nonlinear in $H$ component of $m$ for Dy$_2$O$_3$ paramagnetic salt and sapphire, respectively. The arrows indicate the direction of field changes during these measurements. The compensation method and coefficients $\alpha$ and $\beta$ are the same as described in chapter~\ref{sec:SuperCoil}. (c-f) (dark green bullets) distortions  scans $\eta(z)$ at 5, 20, 40, and 50~kOe with respect to a scan collected at 10~kOe (scaled to the relevant field). Please note that the vertical scale for $\eta(z)$ is in millivolts, not in volts as it is typically for full $V(z)$ indicating an enormous compensation level of $V(z)$. Red lines: $\Upsilon(z)$ fitted to $\eta(z)$ in the linear regression mode. Results of relevant fittings for all the experimental points are marked in red in (a-b). They cross $Y=0$ line at $H = 10$~kOe as this is the measurement that all the points were compensated against. The green lines there are the same as red, but with slightly adjusted slope to level them off at moderate fields. They match then the original $\Delta m$ perfectly, underlying a remarkable linearity of the regression routine.}
   \label{fig:HiFiRes}
\end{figure}

To visualize the effect we present in Fig.~\ref{fig:HiFiRes}(a) compensated for a linear in magnetic field component room temperature data of the same piece of the paramagnetic salt (Dy$_2$O$_3$) as discussed previously in section \ref{sec:SuperCoil}. Here we focus on the sizable outturn of the $\Delta m(H)$ from the expected flat dependency on $H$ seen at above 25~kOe. This nonlinear component in the original $m(H)$ cannot be simply related to deviations from an ideal linearity of this particular paramagnet, as qualitatively the same effect is seen for diamagnetic specimens, as exemplified in Fig.~\ref{fig:HiFiRes}(b). Neither this effect can originate simply from an excess of noise at high fields, since it is seen for both strong and rather weak signals and is smooth and monotonic on $H$. Nor it is caused by the presence of a remnant field in the system, as contrary to $H_{tr}(H)$, the amplitude of the deviations grows with the field and it is at least an order of magnitude stronger than reduction of the magnetic signal due to the presence of $H_{tr}$. Then, the last obvious cause would be a systematically growing uncertainty of the larger magnetic fields in the system. In order to check this possibility we numerically compensate the raw $V(z)$ scans to asses the symmetry of the resulting distortion scan $\eta(z)$. To do so, we take one of the scans from the undistorted part of the $m(H)$ as our reference scan ($V(z)$ obtained at $H=10$~kOe was selected for this purpose). Then we scale this $V(z)$ by the relative value of the magnetic field of each experimental point and subtract such a scaled reference accordingly to obtain sought distortions $\eta(z)$ \footnote{By taking one of the scans obtained during this particular measurement as the reference instead of using the $\Upsilon(z)$ we avoid any uncertainties related to the very details of this particular experimental configuration. These are discussed later in chapters \ref{sec:SampleMounting} and \ref{sec:HandlingArtefacts}}. Four examples of $\eta(z)$ are established this way and are presented in Fig.~\ref{fig:HiFiRes}(c-f). They clearly indicate an odd character of the distortions at large fields and so decisively ruling out any inaccuracies related to a systematic difference between the real magnetic field acting on the sample and the field values reported in the data files.

At the moment we cannot identify the effect responsible for the observed deviations at higher magnetic fields. We have already pointed out (see chapter~\ref{sec:MgnAni}) that an odd component in $V(z)$ is a fingerprint of a presence of a somewhat tilted magnetic moment. Although very likely at weak magnetic fields, here, the moment-tilting should be ruled out based on the increases of the directional uniformity of the field generated in typical solenoids with increasing current in the magnet. Even if the samples would possess any magnetic anisotropy, it should neither be visible at strong fields nor increase with the field. However, whatever the underlying reason, the presence of this odd $\eta(z)$ that causes the least-squares regression routine to be best satisfied for slightly lower values of $m$ than it would be for a 'clean', a perfectly $\Upsilon(z)$-like $V(z)$. This can be confirmed by fitting $\Upsilon(z)$ to the distortion $\eta(z)$. The fitted $\Upsilon(z)$ are marked in panels (c-f) and the corresponding moments (for all the experimental points) are plotted in Figs.~\ref{fig:HiFiRes}(a-b) in red. Their different slope is due to differently levered compensation (giving $\Delta m=0$ at $H=10$~kOe this time), but after re-adjusting the slope the moments obtained from fitting $\eta(z)$ match the original results nearly perfectly (green lines) \footnote{We remark here that in order to mimic the way the original $m(H)$ was established by MPMS, we limit the least-squares regression to so called linear mode in which the fitting routine assumes the central position of the sample in the scanned window. This indeed would be so, as the full data scan $V(z)$ is dominated by the (properly centered) signal of the sample, and so in both linear and iterative modes the $\Upsilon(z)$ is pinned very close to the center.}. As the numbers in Fig.~\ref{fig:HiFiRes} indicate, this 'property' effectively precludes reliable determination of the third significant digit of the magnetic moment by the MPMS at fields stronger than 25~kOe, and so the compensation methods are limited to about 99-99.5\% of compensation of the total signal at high magnetic fields. Finally we observe that despite the fact that the compensation exceeds 99.5\%, the perfect match between values obtained either by a compensation of the results of the fitting (the MPMS results) or by a fit to compensated raw scans (both compensation by a linear term in $H$) indicates a truly remarkable linearity of the regression routine.

\subsubsection{Auto-range option}

It has been reported before that the auto-range option of the SQUID magnetometer can lead to artifacts once the emu-range is changed upon increasing or decreasing total magnetization signals and the characteristic signatures of this type of artifacts have been described in detail before \cite{Ney:2008_JMMM}. Here it should be noted that the emu-range-artifact is not present any more is that particular machine (the MPMS in Duisburg) whereas in the other system in question in the present work, it has never been observed. On the other hand, it was observable in data recorded with other machines as well, e.\,g. in Ref.~\cite{Ney:2007_PRB}, Fig. 1. The disappearance of the emu-range artifact in the MPMS was correlated with the following three actions which unfortunately were carried out at the same time: exchanging of the yellow cables connecting the rf-SQUID-amplifier with the cryostat, re-tuning the SQUID and re-adjusting the filter settings of the rf-amplifier. In any case, this indicates that this particular artifact is indeed correlated with an improper capacitance-match of the R-C-circuits responsible for the various "range"-values as discussed before \cite{Ney:2008_JMMM}. Once such artifacts are noticed, it is therefore advisable to re-tune the SQUID and to check the filter settings. In the worst case, it should be considered to replace the yellow cables. At this point it should be noted that the feed throughs at the cryostat are in some cases not He-leak tight and He blow-off may result in cold plugs and thus condensation of water in these connectors which in turn alters the capacitance of the cables. An "`easy fix"' is to fill the connectors with vacuum grease. Unplugging the yellow cables and drying them out with a heat gun is less advisable, since the connectors are specified for only 40 cycles of plugging and unplugging.

\subsection{Sample mounting}
\label{sec:SampleMounting}

The (second order) gradiometer configuration of the SQUID detection coils, that is the equal number of turns in each directions arranged as shown in Fig.~\ref{fig:SamplePos}, has got two profound implications on the magnetometry. Firstly, the pick-up assembly is far less sensitive to distant moments than those within the volume of the gradiometer. Secondly, there would be no magnetic signal from an infinitely (= sufficiently compared to the size of the gradiometer) long object, providing it is magnetically uniform, independently whether it moves or not and what kind of magnetic response dominates its properties. Practically, anything of a length of 15-20 cm what allows to fix the sample close to its center and has got an adaptor connecting it with the main (long) sample rod will work for sensitive magnetometry, however, the least 'magnetic', the better. The MPMS XL system manufacturer advices to use clear drinking straws as sample holders. Indeed, they are of generally low magnetic signature (the background-signals usually stay below $10^{-7}$~emu) and provide a great versatility when sample fixing is concerned. Since typical specimens are usually of a size of $3 \div 5 \times 5.5$~mm$^2$, all can be held in place inside the straw without any other means than by clamping them in-between the two walls of the straw as illustrated
by Fig.~\ref{fig:SampleInStraw}a). In this configuration the background of the sample holder is minimal and the magnetic field is applied in the sample plane. To reliably perform in-plane versus out-of-plane measurements it is necessary to have a sample piece which can be freely rotated, e.\,g., by using two wooden sticks as non-magnetic "manipulators", inside the {\it same} straw. For that purpose, a $\sim 4 \times 4$~mm$^2$ sample piece can be clamped into the straw along its diagonal as illustrated by Fig.~\ref{fig:SampleInStraw}b). Both options assure sufficiently well-centered samples with respect to the gradiometer axis if straight straws are used.

\begin{figure}[th]
   \begin{center}
        \includegraphics[width=.9\columnwidth]{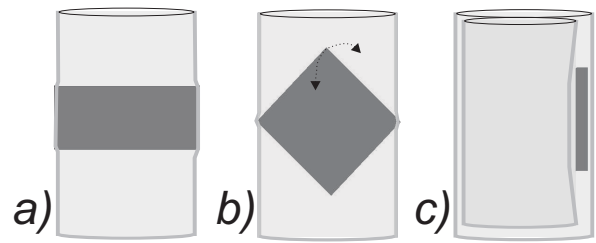}
   \end{center}
   \caption{(Color online) Different means of sample mounting for SQUID measurements using a
straw: a) in-plane measurements for sample sizes $3 - 5 \times 5.5$~mm$^2$,
b) in-plane versus out-of-plane measurements allowing to rotate the sample
($\sim 4 \times 4$~mm$^2$), and c) solution for smaller sample sizes using
a second straw.}
   \label{fig:SampleInStraw}
\end{figure}

In cases where the sample piece is to small to be held freely inside the straw (less than $4 \times 4$~mm$^2$), one has to use a second straw which
has to be carefully slitted along its length ideally using a clean ceramic blade. This second straw can be slightly twisted up while inserted into the
outer straw and presses the sample piece to the wall of the outer straw as illustrated by Fig.~\ref{fig:SampleInStraw}c). The last configuration has to be used in a greater caution when accurate absolute values are desired, as the substantial off-centering of the whole specimen will increase the strength of the coupling with coils (see chapter~\ref{sec:Positioning}). In any case, one has to \emph{strictly} avoid cutting any holes into the walls of the straw, or even severely denting/puncturing the straw in the region of the sample. This inevitably creates an additional (temperature-independent) signal which can be effectively diamagnetic (accumulation of material) or paramagnetic (lack of material).


\begin{figure}[th]
   \begin{center}
        \includegraphics[width=.9\columnwidth]{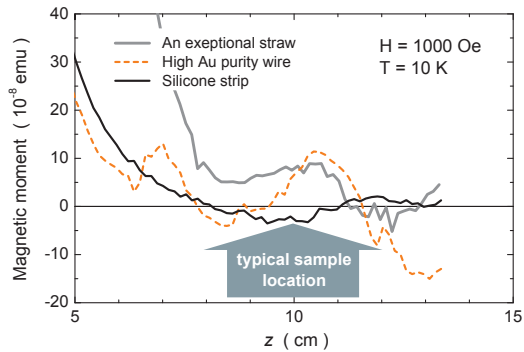}
   \end{center}
   \caption{(Color online) Magnetic moment distribution $\mu(z)$ at the central part of the three different sample holders. The grey line represents one of the very best straw from those provided by Quantum Design, an average straw has got 2-3 times larger amplitude of $\mu(z)$. The dashed orange line represents a typical response of a 0.5 mm gold wire, while the black line gives a typical $\mu(z)$ distribution in 2~mm wide silicon strip. The position $z$ is counted from one of the ends of these sample holders.}
   \label{fig:SHTests}
\end{figure}

Having all the above-mentioned precautions in mind the straws may be of a great value for everyday usage, even precise, magnetometry in a range down to a few $10^{-7}$~emu. For such purposes and to reduce the danger of external magnetic contamination one should avoid re-using straws as well as refrain from cleaning the straws prior to the measurement. In cases where really small signals (in the $10^{-8}$~emu range) are to be researched, one has to look for an alternative. In fact, about 50 straws from two different batches provided by Quantum Design were tested. They were scanned along almost their entire length \footnote{These test were done in a home made SQUID system where scans up to 18 cm in length are possible in an RSO-compatible mode of operation} at 3, 1~kOe and at remanence at 10 and occasionally at 200~K. Then the collected long scans were recalculated to obtain a corresponding (fictitious) moment distribution along their length, the magnetic profile $\mu(z)$. All the test were done without 'front' (top) and the end straw adapters fitted. We wanted to test \textit{the straws} themselves. The presence of straw adapters produces so called "paramagnetic Curie tail" at the lowest temperatures and relatively strong fields \footnote{As this is straw (or a sample holder in general) independent effect we drop this discussion here. There is no way to efficiently mitigate this problem without major surgery into the measurement arrangement of the system.}. We found that the typical amplitude of $\mu(z)$ was a few times $10^{-7}$~emu at 10 K and 1000~Oe, and only in two of all the tested straws $\mu(z)$ was below $10^{-7}$~emu, see an example given in Fig.~\ref{fig:SHTests}. Symptomatically, we found straws from one 'box' to be substantially "less magnetic" (i.e. more magnetically uniform) than from the other. Finally, some proved to be very heavily contaminated, despite the fact that all the straws were thoroughly cleaned in organic solvents prior to their tests. So, to cut on this tedious, time and resources consuming testing job, the Warsaw team started to use a similar length of a straight 4N silver or gold 0.5-0.7~mm wire. They exhibit as a rule the same small $\mu(z)$ as the best straws found by us (dashed orange line in Fig.~\ref{fig:SHTests}), are easy to maintain in 'pristine' condition for a very long time. There are, however, two disadvantages related to this kind of a sample holders. Obviously, a glue is needed to fixed the sample, secondly, the wires deform or bent easily. The last issue has been solved by switching to 1 to 2~mm wide silicon strips professionally cut from 8" commercial wafers \cite{WVRoy}. They are fixed between the standard straw adapters and so are easy to assemble and provide mechanical stability during the measurements. Our tests show that all silicon-based sample holders conform to our stringent requirements concerning magnetic uniformity as we find their $\mu(z)$ to be practically equal to the noise level of the measurement, see black line in Fig.~\ref{fig:SHTests}. However, the major drawback is again the necessity to using a glue to fix the sample. After testing many substances for a possible contribution to the measurement as well as for mechanical reliability and low temperature behavior we settled for the new branch of the GE varnish  \cite{GE_varnish} (heavily diluted), which proved to best fulfill all requirements. It must be finally noted that the usage of Si strips or similar bodies may lead to a sample off centering, so it should be taken into account in very precise magnetometry, see chapter \ref{sec:Positioning}.

\section{Artifacts due to improper sample handling}
\label{sec:HandlingArtefacts}

It is needless to say that every object of any investigations should be as well defined as possible and free from interfering alien contaminants. This is particularly true in integral (SQUID) magnetometry where the typical systems integrate the magnetic response from the specimen and its vicinity. The issue has become of a prime importance how the vast part of the community switched to magnetic investigations of nanoscale magnetic objects and this is there where the main advantage of SQUID magnetometers, their enormous sensitivity, has become also the main concern. This is clearly seen when one realizes that a single 1~ng weighting 5~$\mu$m swarf of iron will produce a moment of  ~$2 \cdot 10^{-7}$~emu, and a typical flake of rust floating in the air around can be easily a few times larger. Thus, the presence of a virtually undetectable (even using a decent optical microscope) contaminations  can totally overwhelm the actual response of a small-signal sample. Some contaminations produce a response which is indistinguishable from that of the sample, so a great care must be exerted at every experimental step, starting from the choice of a substrate, through the deposition and processing until the final mounting on the sample holder. But this is usually not enough, so a very extensive testing augmented by some experience in the field is required to trustworthy separate the real signal of interest (if any) from spurious effects. However, occasionally, contaminations may reveal themselves through a unique, even bizarre, response, effectively very soundly alarming the experienced investigator. We believe that it is very instructive to consider one of such cases in greater detail.

\begin{figure}[th]
   \begin{center}
        \includegraphics[width=0.9\columnwidth]{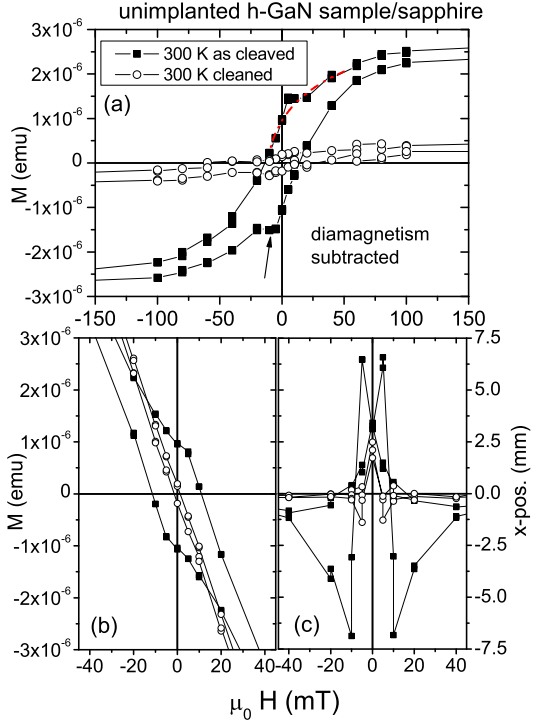}
   \end{center}
   \caption{(Color online) Hysteresis loop of an as-grown GaN layer on sapphire(0001) recorded at 300~K in the as-cleaved state (full symbols) and after cleaning in HCl (open symbols). (a) shows the data after the diamagnetism has been subtracted, (b) the low field part of the full signal, that is with the diamagnetic signal of the substrate, and (c) shows the respective fitted $z$-position of the SQUID signal (iterative regression mode).}
   \label{SampleHandling}
\end{figure}

Such a "fortunate" case is exemplified in Fig.~\ref{SampleHandling} which presents results of the most obvious and presumably common mistake made during the preparation of a sample for SQUID measurements, namely an edge contamination by a (ferrous) material stemming from tweezers or another tool used to prepare, cut, or handle the specimen \cite{Abraham:2005_APL,Garcia:2009_JAP}. This issue has become of a great importance when many laboratories started to investigate materials deposited on substrates made from some of hardest substances like sapphire or SiC which are very likely to abrade the softer material of the tools they are handled by. But even Si requires a great deal of care because of its strong tendency to bind iron or nickel \cite{Abraham:2005_APL,Koo:1995_PRB}, or to absorb on its surface submicron Fe-rich particles during etching from even a high grade KOH \cite{Grace:2009_AM}. Here we consider the case of a sapphire substrate and in Fig.~\ref{SampleHandling}(a) an $m(H)$ measurement at 300~K of an as-cleaved GaN layer on a sapphire(0001) substrate is shown after the diamagnetic background has been subtracted. A clear ferromagnetic-like hysteresis is visible (full squares) with one unusual but alerting feature, namely a kink seen between 10~mT and 25~mT. The first important hint to unravel its origin comes from the inspection of the full moment of the sample, the uncompensated $m(H)$, presented in Fig.~\ref{SampleHandling}(b). We notice that the kink is located in the field range where the raw data changes sign, that is around the compensation field $H_C$ for which values the diamagnetic background become comparable to that of the "ferromagnetic" signal. The second and the most important clue comes from Fig.~\ref{SampleHandling}(c) which shows the fitted "best" $z$-positions of the individual SQUID measurements obtained in the iterative regression mode of the MPMS. It is obvious that the kink corresponds to data points where the fit returns an apparent sample position which is up to 7~mm shifted with respect to the actual position. This can be understood by considering the following: the ferromagnetic signal originates from the edge of the sample at the top (or bottom), whereas the diamagnetic signal comes from the entire sample, i.\,e., it stems from the center of mass of the sample. This experimental situation is depicted in Fig.~\ref{fig:EdgeContaminant}(b). At magnetic fields where both signals are of approx.\ the same size, the respective curves recorded by the MPMS have an \textsl{N}-shaped appearance rather than the expected field distribution with a clear extremum in the center. Thus, the fitting routine picks one of the two extrema of the \textsl{N}-shaped curve and returns a position which is off-center. It is characteristic that this off-set changes sign because the fitting routine takes the extremum of the \textsl{N}-shaped curve which is initially bigger. If the other extremum of the \textsl{N}-shaped curve becomes larger at a different field, the fitting jumps to the other extremum of different sign, therefore causing the characteristic change of sign in the $z$-position plot as shown in Fig.\ \ref{SampleHandling}(c). Characteristically, at strong fields the $z$-position goes to zero, as the properly centered diamagnetic signal dominates, whereas at $H=0$ $z \simeq 3~$~mm, thus very close to the edge of the sample where the contaminant was. And in both limits the shape of the measured $V(z)$ should be very much like the ideal one, which contrasts the case of the tilted magnetic moments (see chapter \ref{sec:MgnAni}) where neither the iterative nor the linear mode of fitting will be acceptable. One may trace the regression fit value for this (reported by the system in .dat files), we would rather advice to inspect the raw data to visually asses each 'problematic' case.

The occurrence of a little kink in the hysteresis at zero measured total magnetization together with the characteristic up-down behavior of the $z$-position in the iterative regression mode as shown in Fig.~\ref{SampleHandling} is thus a clear indication for ferromagnetic contaminants at the edge of the sample. Note that these features would be not visible in the data, if the magnetic contamination is not located at the top or bottom but at the left or right edge (close to the center) with respect to the movement direction of the sample. The safest procedure is to measure $m(H)$ curves with two by 90$^{\circ}$ rotated in-plane configurations. A comparable artefact occurs if a non-homogeneous ferromagnetic sample on a diamagnetic substrate is investigated, e.\,g., nanoparticles deposited at one side of the substrate, ion implantation of only half of the sample or lithographically prepared structures with uneven distribution. Furthermore, if one uses the linear regression mode of the MPMS -- where only the amplitude but not the position is fitted --  one would lose any experimental evidence for magnetic contaminations from the edges of the sample and the data in Fig.\ \ref{SampleHandling}(a) would roughly follow the red line. That the apparent ferromagnetic-like signature in Fig.~\ref{SampleHandling} is indeed caused by a ferromagnetic contamination of the edges of the sample is confirmed by etching the sample in HCl followed by thoroughly cleaning the sample in acetone, ethanol and ultrapure water in an ultrasonic bath for 5~min. each. The $M(H)$ curves were measured once again (open circles) and it is obvious that the magnetic hysteresis has almost vanished and the kink has disappeared (the marginal opening originates from not compensating the data for the presence of the trapped field in the magnet, see chapter~\ref{sec:SuperCoil}). The up-down behavior of the fitted position is now found at zero magnetic field, where the total signal is zero and it is much less pronounced. The overall magnetic behavior is now purely diamagnetic. Furthermore, the role of the GaN layer can be assessed by comparing these findings to the ones in Ref.~\cite{Ney:2008_JMMM}, where a similar behavior has been reported for bare sapphire.

The magnetic contamination at the edge of the sample most likely originates from cleaving it into suitable pieces. It is very common to use a diamond stencil for that purpose; however, the diamond is usually a tiny piece attached to a stainless-steel rod. If this rod touches the edge of a very hard substrate material it is very likely that small particles of the metal are abraded. Hence a thorough cleaning of the sample, especially if sapphire or SiC is used is crucial. In case of substrates which are much softer than stainless steel, the situation is less critical and cleaving usually does not produce any detectable ferromagnetic contamination. Of course no stainless steel tweezers should be used, see Ref.~\cite{Abraham:2005_APL} in any case. Teflon, Titanium or Zirconia tweezers are suitable but one has to assure that they are not used, e.\,g., for nanoparticle samples, in-between since the tweezers-handling always transfers small amounts of nanoparticles to the edges of the sample.

\begin{figure}[h]
   \begin{center}
        \includegraphics[width=0.9\columnwidth]{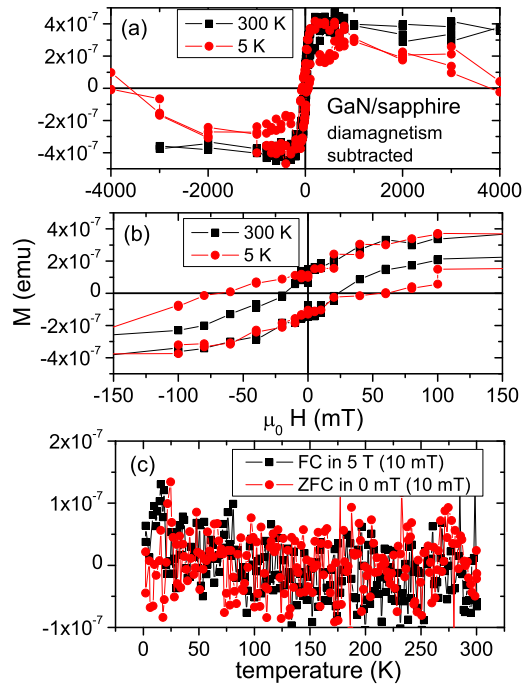}
   \end{center}
   \caption{(Color online) (a) Residual magnetic hysteresis of a GaN layer on sapphire(0001) substrate at 300~K (squares) and 5~K (circles). No additional paramagnetic signal can be seen at low temperatures and high magnetic fields. (b) The apparent hysteresis is not temperature dependent and field cooled (squares) versus zero field cooled (circles) curves coincide indicating the absence long range ferromagnetic order or blocked superparamagnetic behavior (c).}
   \label{residualSignal}
\end{figure}

Once great care has been taken to avoid magnetic contamination of the sample, a sequence can be performed to asses the intrinsic performance of the SQUID for an "every-day" measurement. For that purpose and to "simulate" straightforeward standard measurements without tedious precautions and correction procedures, the identical sample (including straw) as in Fig.\ \ref{SampleHandling} has been further measured at low temperatures following a standard sequence of recording an $m(H)$-loop at 300~K, then at 5~K and then two $m(T)$ measurements while warming under field cooled (FC) and subsequently after zero-field cooled (ZFC) conditions. The outcome is plotted in Fig.~\ref{residualSignal} after subtracting the diamagnetic background which was determined from the mid-field (6 to 10~kOe) $m(H)$-behavior at 300~K, the most simple background correction possible. Figure~\ref{residualSignal}(a) shows two hysteresis curves measured at 300~K and 5~K, revealing virtually identical magnetic response at low temperatures and high magnetic fields with a "`saturation"' magnetization of $4 \cdot 10^{-7}$~emu. At low fields the residual hysteresis becomes visible at 300~K (b). A "remanence" of $2 \cdot 10^{-7}$~emu and a rounded loop with  "coercivity" is visible as well. At 5~K roughly the same "remanence" and "coercivity" is visible. These small residual signals are very similar to the findings in Fig.\ \ref{fig:TrappedField}(a). Since the GaN/sapphire sample is net-diamagnetic as well, it is highly likely that the findings are dominated by $H_{tr}$ in the superconducting magnet. The absence of a true ferromagnetic signal for this sample can be independently inferred by the FC/ZFC curves in Fig.~\ref{residualSignal}(c) which are not separated at any temperature within the noise level and therefore are indicative of the absence of any ferromagnetic or blocked superparamagentic response of the sample. Note, that the $m(T)$ measurements were performed while sweeping (no stabilization of the temperature for each data point) $T$ at a high rate (7~K/min) with little statistics for each point (1 scan, 3 cycles). Even under these "everyday" conditions the residual artifacts as discussed here never exceed $4 \cdot 10^{-7}$~emu and relative changes of the order of $1 \cdot 10^{-7}$~emu can be resolved. However, this is only true, if proper centering according to Fig.\ \ref{fig:SampleInStraw}(a) is used and magnetic contamination is ruled out. For users with little access-time to SQUID magnetometry this may be the ultimate performance achievable and reliable detection of signals below $4 \cdot 10^{-7}$~emu require much more time, care and efforts. Nevertheless, this "realistic" sensitivity still makes the MPMS one of the most sensitive commercial magnetometers.

\section{Conclusion}

In conclusion our detailed analysis of artifacts and limitations of commercial MPMS SQUID magnetometers demonstrate that reaching the intrinsic sensitivity of about $5 \cdot 10^{-9}$~emu (according to the manufacturer's specification) is very challenging if not impossible for samples of finite size, where various magnetic constituents sum up to an overall magnetic response of the specimen. Even for magnetically homogeneous samples, reaching the ultimate sensitivity is difficult when tilted moments occur or the sample is slightly off-axis. In addition, the are artifacts which originate from the intrinsic principle-of-operation of this gradiometric setup. Finally, the use of a superconducting coil to generate the magnetic field adds more complexity, especially for magnetometry at low magnetic fields. Where possible, we have tried to provide typical signatures, concepts to circumvent problems or to compensate for artifacts.

\section*{Acknowledgments}
The authors thank Jim O'Brien of Quantum Design for helpful discussions and comments on the presented here topics. M.S. and W.S acknowledge support by FunDMS Advanced Grant of ERC within the "Ideas" 7th FP of EC, InTechFun (Grant No. POIG.01.03.01-00-159/08), and SemiSpinNet (Grant No. PITNGA-2008-215368). A.N. gratefully acknowledges financial supported from the German Research Foundation (DFG) within the Heisenberg-Programm.

\bibliography{ref_26_12_10}

\end{document}